\documentclass{llncs}

\usepackage{tikzHA}
\usepackage{amsmath,amsfonts,amssymb}
\usepackage{xypic,epsf}
\usepackage{colortbl}
\usepackage{enumitem}
\usepackage{graphicx}
\usepackage{subfigure}
\usepackage{floatflt}
\usepackage{wrapfig}
\usepackage{enumerate}
\usepackage{algorithm}
\usepackage{algorithmic}
\usepackage{import}
\usepackage{listings}
\usepackage{tikz}

\usetikzlibrary{matrix,calc,automata,arrows,arrows.meta,backgrounds,positioning,fit}

\tikzstyle{round state}=[circle,draw,solid,fill=white,line width=1pt]
\tikzstyle{square state}=[rectangle,rounded corners,draw,solid,fill=white,line width=1pt]
\tikzstyle{every node}+=[align=center]
\tikzstyle{every picture}+=[remember picture]
\tikzset{accepting/.style={double distance=2pt},%
every initial by arrow/.style={-{Straight Barb[line width=2pt,length=3mm,width=5mm]}}, %
initial text=,initial distance=1mm,->,line width=0pt,>=stealth',
shorten <= 1pt,shorten >=1pt,auto,node distance=2.8cm, 
every edge/.style={draw,line width=1pt}}

\setenumerate[1]{label={(\arabic*)},leftmargin=*,widest=9}

\newsavebox{\mybox}

\usepackage{etoolbox}
\newcommand{\memberfield}[3]{{%
   \ifstrequal{.}{#1}{%
     {\sf \dot{#3}}{\ifx&#2&{}\else (#2)\fi}%
     }{%
     {\sf #3 #1}{\ifx&#2&{}\else (#2)\fi}%
     }
   }}

\newcommand{\pointer}[2][]{{\memberfield{#1}{#2}{p}}}
\newcommand{\variable}[2][]{{\memberfield{#1}{#2}{x}}}
\newcommand{\pos}[2][]{{\memberfield{#1}{#2}{pos}}}

\newcommand{\lane}[2][]{{\memberfield{#1}{#2}{lane}}}
\newcommand{\back}[2][]{{\memberfield{#1}{#2}{back}}}
\newcommand{\front}[2][]{{\memberfield{#1}{#2}{front}}}

\newcommand{\sideback}[2][]{{\memberfield{#1}{#2}{sideback}}}
\newcommand{\sidefront}[2][]{{\memberfield{#1}{#2}{sidefront}}}

\newcommand{\nil}{{\sf nil}}

\newcommand{\ignore}[1]{}

\newcommand{\environment}{\text{\sf Top}}

\newtheorem{thm}{Theorem}

\newtheorem{cor}[thm]{Corollary}
\newtheorem{defi}{Definition}

\newtheorem{ex}{Example}

\newcommand{\flow}{{\sf flow}}
\newcommand{\invariant}{{\sf Inv}}
\newcommand{\initialStates}{{\sf Init}}
\newcommand{\guard}{{\sf guard}}
\newcommand{\jump}{{\sf jump}}

\newcommand{\K}{{\mathcal K}}
\newcommand{\T}{{\mathcal T}}

\long\def\ignore#1{}

\begin{document}
\title{On verification and constraint generation for families of
  similar hybrid automata}
\author{Viorica Sofronie-Stokkermans \and
Philipp Marohn}
\institute{University of Koblenz, Germany}

\maketitle

\vspace{-2mm}
\begin{abstract}
In this paper we give an overview of results on the analysis of 
parametric linear hybrid automata, and of systems of similar linear 
hybrid automata: We present possibilities of describing systems with a parametric 
(i.e.\ not explicitly specified) number of similar components which
can be connected to other systems, such that some parts in the
description might be underspecified (i.e.\ parametric). We consider 
global safety properties for such systems, expressed by universally 
quantified formulae, using quantification over variables
ranging over the component systems. 
We analyze possibilities of using methods for hierarchical reasoning and 
symbol elimination for determining relationships on (some of) the 
parameters used in the description of these systems under which the 
global safety properties are guaranteed to be inductive invariants. 
We discuss an implementation and illustrate its use on several 
examples. 
\end{abstract}

\vspace{-2mm}
\section{Introduction}
In this paper we give an overview of some of our results on the analysis of 
systems of parametric linear hybrid automata, with a focus on
identifying possibilities of generating constraints on parameters
under which given safety properties are guaranteed to hold, and
illustrate the way we used an implementation of a method for symbol 
elimination in theory extensions for solving such problems.

A considerable amount of work has been dedicated in the past to identifying 
classes of hybrid automata for which checking safety is decidable. 
While reachability and safety in linear hybrid automata are 
in general undecidable, invariant checking and bounded reachability are decidable.
There exist approaches to the verification of parametric reactive infinite 
state systems and timed automata 
(e.g.\ by Ghilardi et al. \cite{Ghilardi}, Hune et
al. \cite{HuneRomijn}, 
Cimatti et al.\ \cite{Cimatti2}) and for parametric hybrid automata (e.g.\
by Henzinger et al. \cite{Henzinger}, 
Frehse \cite{FrehseJK08}, Wang \cite{farn-wang}, Cimatti et 
al.\ \cite{Cimatti}, Fr{\"a}nzle \cite{FranzleGKAK15} (for probabilistic hybrid
systems)) but in most cases only situations in which the 
parameters are constants were considered. 
In this context we also mention the development and study of a dynamic
hybrid logic \cite{Platzer08,Platzer12-lics,CordwellPlatzer-19cade}, 
as well as existing tools 
(cf.\ e.g.\ \cite{FrehseJK08,Frehse-13,FribourgK13,keymaera-16,keymaera-20,keymaera-22-ijcar}). 
Systems of systems have been studied in many papers 
(cf. e.g.\
\cite{sofronie-entcs,Faber-Jacobs-Sofronie-07,FaberIJS10,KaiserKW10,FreseB10,HilscherLOR11,Platzer-12-lmcs,JohnsonMitra12b,JohnsonMitra12,AbdullaHH13,DammPRW13,HenzingerMineaPrabhu,DammHS15,JaberJW0S20,KrogerF22,JaberWJKS23}
to mention only a few, cf. also \cite{JacobsBook} for further references). 
Many such papers prove small model or cutoff properties. 

We analyzed possibilities of using hierarchical reasoning for the
verification of linear hybrid systems and of systems of hybrid systems
in \cite{Sofronie-Stokkermans10,DammIS11,Sofronie-cade2013,DammHS15,Sofronie-Fundamenta-Informaticae-2020}. 
The results presented in this paper are based on \cite{DammHS15}, in
which a definition of systems of hybrid automata is proposed. 
In \cite{Sofronie-cade2013,DammIS11,DammHS15}
we showed that methods for hierarchical reasoning in complex theories
can be used to identify classes of (systems of) hybrid automata for which decision procedures
exist, but also for deriving additional assumptions on the properties of
  parameters which guarantee that a certain safety property is an
  invariant. In the tests presented in \cite{DammIS11,DammHS15} we only considered 
the problem of checking whether given formulae were inductive
invariants, and some of the constants were replaced with concrete numbers in
order to generate linear constraints. Since we used as an backend solver the
version of Z3 available at that time, checking validity of non-linear
constraints was problematic. 
The results we present here bridge this gap: we
consider parametric problems and use 
quantifier elimination for generating constraints on parameters. 
We present a way of describing such 
systems proposed in \cite{DammHS15} and discuss a method for determining 
relationships on (some of) the parameters used in the description 
of these families of systems based on symbol elimination and its implementation in the
system SEH-PILoT. We then illustrate the way 
SEH-PILoT can be used for constraint generation. 

Our work in this area was greatly influenced by the collaboration in
the AVACS project in general and  by the fruitful discussions with 
Martin Fr{\"a}nzle in particular. We therefore dedicate this paper to him.

\smallskip
\noindent 
{\em Structure of the paper.}
In Section~\ref{idea} we present
some 
examples which illustrate the problems we consider. 
In Section~\ref{Sec:Preliminaries} we introduce the notions in logic
and in Section~\ref{Sec:HierarchicalReasoning} the notions
on hierarchical reasoning needed in the paper. 
In Section~\ref{Sec:HybridAutomata} we introduce hybrid automata and 
linear hybrid automata and the
verification problems we consider. In Section~\ref{Sec:FamiliesLHA} we present
the way we defined systems of similar hybrid automata in \cite{DammHS15}
and the related verification problems, and give some examples which
show how constraints on parameters which guarantee safety in systems of
linear hybrid automata can be automatically generated. 
In Section~\ref{Sec:Conclusions} we present some conclusions and plans
for future work. 

\vspace{-2mm} \subsection*{Table of Contents}

\contentsline {section}{\numberline {1}Introduction}{1}{}%
\contentsline {section}{\numberline {2}Idea}{3}{}%
\contentsline {section}{\numberline {3}Preliminaries}{5}{}%
\contentsline {section}{\numberline {4}Local theory extensions}{5}{}%
\contentsline {subsection}{\numberline {4.1}Examples of local theory extensions}{6}{}%
\contentsline {subsection}{\numberline {4.2}Symbol elimination in local theory extensions}{8}{}%
\contentsline {subsection}{\numberline {4.3}Tools}{10}{}%
\contentsline {section}{\numberline {5}Parametric Linear Hybrid Automata}{11}{}%
\contentsline {subsection}{\numberline {5.1}Verification}{13}{}%
\contentsline {subsection}{\numberline {5.2}Example: Verification and constraint generation}{14}{}%
\contentsline {section}{\numberline {6}Families of Similar Hybrid Automata}{18}{}%
\contentsline {subsection}{\numberline {6.1}Verification}{19}{}%
\contentsline {subsection}{\numberline {6.2}Examples: Constraint generation}{20}{}%
\contentsline {section}{\numberline {7}Conclusions}{30}{}%

\section{Idea}
\label{idea}
We illustrate the problems studied in the paper
 on the following examples: 

\begin{ex}[\cite{Sofronie-Fundamenta-Informaticae-2020}]
\label{ex1}
{\em 
We can model a water tank controller as a hybrid system, with
variable $L$ (water level) and two modes $s_1, s_2$ (state invariants 
$L \geq L_{\sf a}$ and $L < L_{\sf a}$, where $L_{\sf a}$ is an alarm
level). In mode $s_1$ we have inflow and outflow of water; in mode
$s_2$ only inflow.  
The water level, as well as the inflow and outflow are modeled using
unary functions $L$, ${\sf infl}, {\sf outfl}$, where
$L(t)$, ${\sf infl}(t)$ and ${\sf outfl}(t)$ are the water level,  the
inflow and  outflow at time $t$, respectively. 
We here assume that the inflow and outflow rates are constant 
and equal to ${\sf in}$, resp.\ ${\sf out}$ (i.e.\ the derivative of
${\sf infl}$ is equal to ${\sf in}$ at every point in time $t$ and the derivative of
${\sf outfl}$ is equal to ${\sf out}$ at every point in time $t$). 
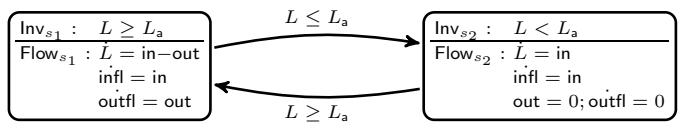
\begin{center}
{\scriptsize 
\begin{tikzpicture}
\node [square state] (1) {$\begin{array}{ll}
{\sf Inv}_{s_1}: \quad & L \geq L_{\sf a}\\
\hline 
{\sf Flow}_{s_1}: & \dot{L} = {\sf in} {-} {\sf out}\\
& \dot{{\sf infl}} = {\sf in}\\
& \dot{{\sf outfl}} = {\sf out}
\end{array}$}; 
\node [square state] (2) [right = of 1]{$\begin{array}{ll}
{\sf Inv}_{s_2}: \quad  & L <
L_{\sf a}\\
\hline 
{\sf Flow}_{s_2}: & \dot{L} = {\sf in}\\
& \dot{{\sf infl}} = {\sf in}\\
& {\sf out} = 0; \dot{{\sf
    outfl}} = 0
\end{array}$};
 
\path 
(1) edge [bend left=10,above] node  {$L \leq L_{\sf a}$} (2)
(2) edge [bend left=10,below] node  {$L \geq L_{\sf a}$} (1);  
\end{tikzpicture}
}
\end{center}

\noindent Clearly, after an evolution from time $t_0$ to time $t_1 > t_0$ in mode $s_1$ (resp. $s_2$) 
we have $L(t_1)  = L(t_0) + ({\sf in} - {\sf out})*(t_1 - t_0)$ (resp.\ $L(t_1) = L(t_0) +
{\sf in}*(t_1 - t_0)$).

Consider the safety condition $\Psi = L \leq L_{\sf o}$ stating that
the water level always remains below an overflow level, $L_{\sf o}$. 
Since in the mode changes $L$ is not updated, $\Psi$ is clearly
invariant under jumps. 
$L \leq L_{\sf o}$ is invariant under flows iff the following
formulae are unsatisfiable w.r.t.\ the theory ${\cal T}_S$ of real numbers: 
\begin{itemize}
\item[(i)]  $\exists L, t (L {\leq} L_{\sf o} \wedge 0 {<} t
  \wedge L {\geq} L_{\sf a} \wedge \forall t'( 0 {\leq} t'
  {\leq} t {\rightarrow} L {+} {\sf in}'\!{*}t' \geq L_{\sf a})
  \wedge L {+} {\sf in}'\!{*}t {>} L_{\sf o})$,
\item[(ii)] $\exists L, t (L {\leq} L_{\sf o} \wedge 0 {<} t \wedge L {<} L_{\sf a} \wedge \forall t' (0 {\leq} t' {\leq} t {\rightarrow} L {+} {\sf in}{*}t' < L_{\sf a}) \wedge L {+} {\sf in}{*}t {>} L_{\sf o})$,
\end{itemize}
where in (i) ${\sf in}'$ is used as an abbreviation for ${\sf in} - {\sf
  out}$. 
In \cite{DammIS11} we proved that (i) and (ii) are unsatisfiable iff
(i') and (ii') are unsatisfiable: 
\begin{itemize}
\item[(i')] $\exists L, t (L {\leq} L_{\sf o} \wedge 0 {<} t
  \wedge L {\geq} L_{\sf a} \wedge  L {+} ({\sf in} {-} {\sf out}) {*}t \geq L_{\sf a}
  \wedge L {+} ({\sf in}{-}{\sf out}){*}t {>} L_{\sf o})$,
\item[(ii')]  $\exists L, t (L {\leq} L_{\sf o} \wedge 0 {<} t \wedge L {<} L_{\sf a} \wedge  L {+} {\sf in}{*}t < L_{\sf a} \wedge L {+} {\sf in}{*}t {>} L_{\sf o})$.
\end{itemize}
We can use quantifier elimination in the theory of real closed fields 
to obtain weakest constraints on the parameters ${\sf in}$ and ${\sf out}$
under which (i') and (ii') are unsatisfiable.
}
\end{ex}

\begin{ex} 
\label{ex2}
{\em 
Consider a family of $n$ water tanks with a uniform description, 
each modeled by a hybrid automaton $S(i)$ similar to the one described in Example~\ref{ex1}. 
Assume that for every $S(i)$ the water level in the tank is
represented by the continuous variable $L(i)$, and that the rate of
inflow and outflow for system $S(i)$ are constants depending on $i$, 
and are described by parameters 
${\sf in}(i)$ and ${\sf out}(i)$. 
Assume that, for every $i$, ${\sf in}(i) \geq 0$ and ${\sf out}(i) \geq 0$. 
Assume that the water tanks are interconnected in such a way that 
the output of system $S(i)$ is the input of system $S(i+1)$.
Our goal is to automatically obtain a weakest
universal condition on the parameters 
which guarantees that the formula 
$\forall i (L(i) \leq L_{\sf o})$, where $L_{\sf o}$ is an overflow level, is an
inductive invariant of this system of hybrid automata. 
For this we need a way of eliminating also function symbols.
}
\label{ex-water-tanks}
\end{ex}
The way the systems in Example~\ref{ex-water-tanks} 
are interconnected does not change in time. The next example 
refers to a situation in which the interconnections between systems 
might change. 
\begin{ex}[\cite{DammHS15}]
{\em We consider a family of similar (but not identical) autonomous cars on a highway. 
A car can observe other cars through sensors. 
If the highway has one lane, every car should be able to observe the
closest car in front and possibly also in the back. 
In \cite{DammHS15} we considered highways with
two lanes;  for describing the
closest car in front, back, in the front on the other lane and in the
back on the other lane we use unary
functions
 $\back{}$, $\front{}$, $\sidefront{}$, $\sideback{}$. 
We assume that the behavior of the cars $i \in I$ is controlled by similar hybrid
automata $S(i), i \in I$ such that for each $i$ the automaton $S(i)$ has
two modes: one mode in which the car $i$ tries to reduce the
distance to the car in front of it (${\sf front}(i)$) because the
distance between them is above a certain value 
$d_{\sf appr}$ and one mode in which car $i$ 
tries to increase the distance to the car in front of it (${\sf front}(i)$) because the
distance between them is below a certain value
$d_{\sf rec}$. 
The topology of the system can change: The cars can change their lane, 
and in fixed intervals of time, the links $\back{}$, $\front{}$,
$\sidefront{}$, $\sideback{}$
are updated depending on the actual positions of the cars. 
A verification task we considered in \cite{DammHS15} was to check whether the
distance between a car and the car in front of it on the same lane is
always larger than a safety distance $d_{\sf safe}$. 
If this is not possible, it might be useful  to  
obtain constraints on the functional parameters $d_{\sf appr}$, $d_{\sf
  rec}$ and $d_{\sf safe}$ such that this is guaranteed. 
}\end{ex}

\section{Preliminaries}
\label{Sec:Preliminaries}

We present the notions in logic needed in this paper.

\medskip
\noindent {\bf Logic.} We consider signatures of the form
$\Pi = (\Sigma, {\sf Pred})$ or many-sorted signatures of the form 
$\Pi = (S, \Sigma, {\sf Pred})$, 
where $S$ is a set of sorts, $\Sigma$ is a family of function symbols and ${\sf Pred}$
a family of predicate symbols. 
If $\Pi$ is a signature and $C$ is a set of new constants, we will denote 
by $\Pi^C$ the expansion of $\Pi$ with constants in $C$, i.e.\ 
the signature $\Pi^C = (\Sigma \cup C, {\sf Pred})$. 
We assume known standard definitions from first-order logic  
such as terms, atoms, formulae, $\Pi$-structures, logical entailment,  
model, satisfiability, unsatisfiability.  
A literal is an atom or the negation of an atom; a clause is a
(finite) disjunction of literals. In this paper we refer to (finite) conjunctions of
clauses also as ``sets of clauses'', and to (finite) conjunctions of formulae 
as ``sets of formulae''. Thus, if $N_1$ and $N_2$ are finite sets of
formula then $N_1 \cup N_2$ 
will stand for the conjunction of all formulae in $N_1 \cup N_2$. 
All free variables of a clause (resp. of a set
of clauses) are considered to be implicitly universally quantified. 
We denote ``verum'' with $\top$ and ``falsum'' with $\perp$. $\perp$ is also a notation 
for the empty clause. 

\medskip
\noindent 
{\bf Logical theories.} 
First-order theories are sets of formulae (closed under logical consequence), 
typically all consequences of a set of axioms. 
Alternatively, we may consider a set of models which defines a
theory.
Theories can be defined by specifying a set of axioms, or by specifying a
set of structures (the models of the theory). 
In this paper, (logical) theories are simply sets of sentences.

\smallskip
\noindent 
If $F, G$ are formulae and ${\mathcal T}$ is a theory we 
write $F \models G$ to express the fact that every model of $F$ is a
  model of $G$ and 
$F \models_{\mathcal T} G$ -- also written as ${\mathcal T} \cup F
  \models G$ and sometimes ${\mathcal T} \wedge F \models G$ -- 
to express the fact that every model of $F$ which is also a model of 
$\T$ is a model of $G$.
If $F \models G$ we say that {\em $F$ entails $G$}. If $F \models_{\mathcal T}
 G$ we say that {\em $F$ entails $G$ w.r.t.\ ${\mathcal T}$}. 
$F \models \perp$ means that $F$ is
unsatisfiable; $F \models_{\T} \perp$ means that there is no model of
$\T$ in which $F$ is true. 
If there is a model of $\T$ which is also a
model of $F$ we say
that $F$ is satisfiable w.r.t.\ ${\mathcal T}$. 
If $F \models_{\mathcal T} G$ and $G \models_{\mathcal T} F$ we say that
 {\em $F$ and $G$ are equivalent w.r.t.\ ${\mathcal T}$}.

\section{Local theory extensions} 
\label{Sec:HierarchicalReasoning}
We now introduce a class of theories for which decidable fragments relevant 
to the tasks we consider exist.

Let ${\cal T}_0$ be a base theory with signature $\Sigma_0$. 
We consider extensions ${\cal T}_1 := {\cal T}_0 \cup {\cal K}$ of ${\cal T}_0$ with new 
function symbols in a set $\Sigma_1$ of {\em extension functions}
whose properties are axiomatized with a set ${\cal K}$ of 
{\em clauses}. 
In this case we refer to the (theory) extension ${\cal T}_0 \subseteq {\cal T}_0 \cup {\cal K}$. 
In \cite{Sofronie-cade-05} we introduced and studied local theory
extensions. In \cite{ihlemann-sofronie-ijcar10}, various notions of
locality of  theory extensions were introduced and studied. We present
some of these definitions and results below. 

\begin{defi}[Local theory extension]
An extension ${\cal T}_0 \subseteq {\cal T}_0 \cup {\cal K}$ is a
{\em local extension} 
if for every set $G$ of 
ground $\Pi^C$-clauses 
(where $C$ is a set of additional constants), 
if $G$ is unsatisfiable w.r.t.\ ${{\cal T}_0 {\cup} {\cal K}}$ then 
unsatisfiability can be detected using the set ${\cal K}[G]$ 
consisting of those instances 
of ${\cal K}$ in which the terms starting with 
extension functions are ground terms occurring in ${\cal K}$ or $G$.
\label{defi-loc}
\end{defi} 
\emph{Stably local} extensions are defined similarly, with the difference
that ${\cal K}[G]$ is replaced with ${\cal K}^{[G]}$, the set of 
instances of ${\cal K}$ in which the variables are instantiated 
with ground terms 
which occur in ${\cal K}$ or $G$.

In \cite{Sofronie-cade-05} we showed that 
local theory extensions can be 
recognized by showing that certain partial models embed into total
ones. If a theory extension has the property that each such partial
model embeds into a total model with the same universe, we talk 
about completability (we express this condition as {\sf Comp}).

\medskip
\noindent {\bf Hierarchical reasoning in local theory extensions.}
For local theory extensions (or stably local theory extensions) 
hierarchical reasoning is possible. If ${\cal T}_0 \cup {\cal K}$ is a
(stably) local extension of ${\cal T}_0$ and $G$ is a set of 
ground $\Pi^C$-clauses 
then, by Definition~\ref{defi-loc}, 
$ {\cal T}_0 \cup {\cal K} \cup G$ is unsatisfiable iff 
$ {\cal T}_0 \cup {\cal K}[G] \cup G$ (or resp.\ 
$ {\cal T}_0 \cup {\cal K}^{[G]} \cup G$) is unsatisfiable. 
We can reduce this last satisfiability test to 
a satisfiability test w.r.t.\ ${\cal T}_0$. 
The idea is to purify 
${\cal K}[G] \cup G$ (resp.\ ${\cal K}^{[G]} \cup G$) 
by 
\begin{itemize}
\vspace{-1mm}
	\item introducing (bottom-up) new constants $c_t$ for subterms $t = f(g_1, \dots, g_n)$ with $f \in \Sigma_1$, $g_i$ ground $\Sigma_0 \cup \Sigma_c$-terms,
  \item replacing the terms $t$ with the constants $c_t$, and 
	\item adding the definitions $c_t = t$ to a set $D$. 
\vspace{-1mm}
\end{itemize}
 We denote by ${\cal K}_0 \cup 
G_0 \cup D$ the set of formulae obtained this way. 
Then $G$ is 
satisfiable w.r.t.\ ${\cal T}_0 \cup {\cal K}$ iff 
${\cal K}_0 \cup G_0 \cup {\sf Con}_0$ is satisfiable w.r.t.\
${\cal T}_0$, where 

\smallskip
${\sf Con}_0  = \{ (\bigwedge_{i = 1}^n c_i {=} d_i) \rightarrow c {=} d \mid 
  f(c_1, \dots, c_n) {=} c, f(d_1, \dots, d_n){=} d \in D \}.$

\begin{thm}[\cite{Sofronie-cade-05}] 
\label{lemma-rel-transl}
If ${\cal T}_0 \subseteq {\cal T}_0 \cup {\cal K}$ is a 
(stably) local extension and $G$ is a set of
 ground clauses 
then we can reduce the problem of checking whether $G$ is 
satisfiable w.r.t.\ ${\cal T}_0 \cup {\cal K}$ to checking 
the satisfiability 
w.r.t.\ ${\cal T}_0$ of the formula ${\cal K}_0 \cup G_0 \cup {\sf
  Con}_0$ constructed as explained above. 
If ${\cal K}_0 \cup G_0 \cup {\sf Con}_0$ belongs to a decidable 
fragment ${\cal F}$ of ${\cal T}_0$ we can use the decision procedure for 
this fragment to decide whether $ {\cal T}_0 \cup {\cal K} \cup G$
is unsatisfiable. 
\end{thm}
As the size of  
${\cal K}_0 {\cup} G_0 {\cup} {\sf Con}_0$ is polynomial in the size of $G$
(for a given ${\cal K}$), locality allows us to express the complexity 
of the ground satisfiability problem w.r.t.\ ${\cal T}_1$  
as a function of the complexity of the satisfiability 
of formulae in ${\cal F}$ w.r.t.\ ${\cal T}_0$. 

\subsection{Examples of local theory extensions}
\label{examples}
In establishing the decidability results for the verification of
safety properties of the systems of linear hybrid automata we
consider, we will use locality results for updates and for
theories of pointers. Below are some of these locality results.

\medskip 
\noindent {\bf Uninterpreted functions:} The extension ${\cal T}_0
\cup {\sf UIF}_{\Sigma}$ of any theory
  ${\cal T}_0$ with a set $\Sigma$ of uninterpreted function 
symbols is local and satisfies condition ${\sf Comp}$.

\smallskip
\noindent {\bf Boundedness \cite{sofronie-ihlemann-ismvl-07,Sofronie-Ihlemann-Jacobs-tacas08}:} Assume ${\cal T}_0$ contains a reflexive 
binary predicate $\leq$, and $f \not\in \Sigma_0$. Let $m \in {\mathbb N}$. 
For $1 \leq i \leq m$ let $t_i(x_1, \dots, x_n)$ and 
$s_i(x_1,\dots, x_n)$ 
be terms in the signature $\Pi_0$ 
and $\phi_i(x_1, \dots, x_n)$ 
be $\Pi_0$-formulae with (free) variables among $x_1, \dots, x_n$, 
such that (${\overline x}$ denotes the sequence $x_1, \dots,
x_n$): 
\begin{itemize}
\vspace{-1mm}
\item[(i)]${\cal T}_0 \models \forall {\overline x} 
(\phi_i({\overline x})\rightarrow s_i({\overline x}) \leq t_i({\overline x}))$, 
and 
\item[(ii)]
if $i \neq j$, $\phi_i \wedge \phi_j \models_{{\cal T}_0} \perp$.
\end{itemize}

\noindent Let ${\sf GB}_f  =  \bigwedge_{i = 1}^m {\sf GB}_f^{\phi_i}$ and  
${\sf Def}_f  =   \bigwedge_{i = 1}^n {\sf Def}_f^{\phi_i}$, where:

\smallskip
\noindent $({\sf GB}_f^{{\phi_i}})~~ \forall {\overline x}
(\phi_i({\overline x}) \rightarrow  s_i({\overline x}) \leq
f({\overline x}) \leq t_i({\overline x})) \quad 
({\sf Def}_f^{{\phi_i}})~~ \forall {\overline x} (\phi_i({\overline x}) \rightarrow  f({\overline x}) = t_i({\overline x}))$

\smallskip
\noindent The extensions 
${\cal T}_0 \subseteq {\cal T}_0 \cup {\sf GB}(f)$ and  ${\cal T}_0 \subseteq {\cal T}_0 \cup {\sf Def}(f)$ are both 
local \cite{sofronie-ihlemann-ismvl-07,Sofronie-Ihlemann-Jacobs-tacas08}.

\smallskip
\noindent 
{\bf Updates \cite{JacobsKuncak,Sofronie-Ihlemann-Jacobs-tacas08}:}
Let ${\cal T}_0$ be a theory with signature $\Sigma_0$ and
$\Sigma \subseteq \Sigma_0$. Let $\Sigma' = \{ f' \mid f \in \Sigma
\}$, where $f'$ represents the value of the function $f$ after the update.
Consider a family ${\sf Update}(\Sigma, \Sigma')$ 
of update axioms of the form: 

\medskip
$\forall {\overline x} (\phi^f_i({\overline x}) \rightarrow
    F^f_i(f'({\overline x}), {\overline x})),  i =1, \dots, m,
    \quad f \in \Sigma$   

\medskip
\noindent which describe how the values of the $\Sigma$-func\-tions change,  
depending on a partition of the state space, described by a finite set $\{ \phi^f_i \mid i \in I \}$ of
$\Sigma_0$-formulae and using $\Sigma_0$-formulae $F^f_i$ such that
\begin{itemize}
\item[(i)] $\phi_i({\overline x}) \wedge \phi_j({\overline x})
  \models_{{\cal T}_0} \perp $  for $i {\neq} j$ and
\item[(ii)] ${\cal T}_0 \models \forall {\overline x} (\phi_i({\overline x})
  \rightarrow \exists y (F_i(y, {\overline x})))$ for all $i \in I$.
\end{itemize}
Then the extension of ${\cal T}_0$ with axioms ${\sf Update}(\Sigma, \Sigma')$
is local. 

\smallskip
\noindent 
{\bf Theory of pointers \cite{NeculaMcPeak,Sofronie-Ihlemann-Jacobs-tacas08}:}
Consider the language ${\cal L}_{{\sf index},{\sf num}}$ 
with sorts ${\sf index}$  and 
${\sf num}$,
with sets of unary pointer fields $P$ with arity ${\sf index}
\rightarrow {\sf index}$ and numeric fields $X$ with arity ${\sf
  index} \rightarrow {\sf num}$, and
with a constant $\nil$ of sort ${\sf index}$.
The only predicate of sort ${\sf index}$ is equality; the signature
$\Sigma_{\sf num}$ 
of sort ${\sf num}$ depends on the theory ${\cal T}_{\sf num}$ modeling the
scalar domain. A {\em guarded {\sf index}-positive extended clause} is 
a clause of the form: 
\begin{eqnarray}
\vspace{-5mm}
\forall i_1 \dots i_n ~~ {\cal E}(i_1, \dots, i_n) \vee {\cal
  C}({\overline x_i(i_1)}, \dots, {\overline x_i(i_n)}) \label{loc-ax}
\vspace{-5mm}
\end{eqnarray}
\noindent where ${\cal C}$ is a ${\cal T}_{\sf num}$-formula over terms
of sort ${\sf num}$, 
$x_i \in X$, and ${\cal E}$ is a disjunction of equalities between
terms of sort ${\sf index}$, containing all atoms of the form 
$i = \nil,  f_n(i) = \nil, \dots, 
f_2(\dots f_n(i)) = \nil$  for all terms $f_1(f_2(\dots f_n(i)))$ occurring in 
${\cal E} \vee {\cal C}$, where $f_1 \in P \cup X, f_2,\dots, f_n \in P$. 

\noindent Every set ${\cal K}$ of 
guarded {\sf index}-positive extended clauses defines a stably local 
extension of ${\cal T}_{\sf num} \cup {\sf Eq}_{\sf index}$, 
where ${\sf Eq}_{\sf index}$ is the pure theory
of equality.
\label{remark-pointers}

\medskip
\noindent Other examples which turned out to be useful in the study of 
parametric systems were e.g.\ theories of monotone functions
\cite{Sofronie-cade-05,sofronie-ihlemann-ismvl-07}
and 
theories of convex and concave functions defined on an interval $I$ of real
numbers or integers \cite{sofronie-ki08}. 

\medskip
\noindent {\bf Chains of local theory extensions.}
In many cases we need to perform reasoning tasks 
in an extension ${\cal T}_0 \subseteq
{\cal T}_0 \cup {\cal K}$ in which the set ${\cal
  K}$ of axioms of the extension can be written as a union 
${\cal  K} = {\cal  K}_1 \cup {\cal  K}_2$ such that

\smallskip
${\cal T}_0 \subseteq
{\cal T}_0 \cup {\cal K}_1$ and 
${\cal T}_0 \cup {\cal K}_1  \subseteq {\cal T}_0 \cup {\cal K}_1 \cup
{\cal K}_2$ 

\smallskip
\noindent are both (stably) local theory
extensions.
In this case we say that we have a chain of (stably) local 
theory extensions; the reasoning task can be hierarchically reduced 
to reasoning in ${\cal T}_0$ in two steps: 
\begin{description}
\item[Step 1:]  In a first step, we reduce checking whether 
${\cal T}_0 \cup {\cal K}_1 \cup {\cal K}_2 \cup G$ is satisfiable
to checking whether ${\cal T}_0 \cup {\cal K}_1 \cup {\cal K}_2*[G]
\cup G$
is satisfiable (where ${\cal K}_2*[G]$ is ${\cal K}_2[G]$ if the
extension is local and ${\cal K}_2^{[G]}$ if it is stably
local). 
We can further reduce this task to a satisfiability task in ${\cal T}_0
\cup {\cal K}_1$
as explained in Theorem~\ref{lemma-rel-transl}. 

\item[Step 2:]  If all variables in $\K_2$ occur below extension
  functions then $G_1 = ({\cal K}_2)_0 \cup G_0 \cup {\sf
  Con}_0$ is a set of ground clauses. If the theory extension 
${\cal T}_0 \subseteq
{\cal T}_0 \cup {\cal K}_1$ is (stably) local, we can again use  
Theorem~\ref{lemma-rel-transl} to reduce the problem of checking the 
satisfiability of 
${\cal T}_0 \cup {\cal K}_1 \cup G_1$ to a satisfiability test w.r.t.\
${\cal T}_0$. 
\end{description}
The idea can be used also for longer chains of (stably) local theory
extensions: 

${\cal T}_0 \subseteq {\cal T}_0 \cup {\cal K}_1 \subseteq {\cal
  T}_0 \cup {\cal K}_1 \cup {\cal K}_2 \subseteq \dots \subseteq  {\cal
  T}_0 \cup {\cal K}_1 \cup {\cal K}_2 \cup \dots \cup {\cal K}_n.$

\subsection{Symbol elimination in local theory extensions}
\label{symbol-elimination}

Let $\Pi_0 = (\Sigma_0, {\sf Pred})$. 
Let ${\mathcal T}_0$ be a base theory with signature $\Pi_0$.  
We consider theory extensions $\T_0
\subseteq \T = \T_0 \cup \K$, in which among the extension functions
we identify a set of {\em parameters} $\Sigma_P$ (function and constant symbols). 
Let $\Sigma$ be a signature consisting of extension symbols which are not
parameters (i.e.\ such that $\Sigma \cap (\Sigma_0 \cup \Sigma_P) =
\emptyset$). Let $\Pi = (\Sigma_0 \cup \Sigma_P \cup \Sigma, {\sf Pred})$.

\medskip
\begin{algorithm}[h!]
\caption{Algorithm for Symbol Elimination in Theory Extensions \cite{sofronie-ijcar2016,sofronie-lmcs-2018}}

\begin{tabular}{ll}
{\bf Input:} & \!\!\!\!Theory extension ${\cal T}_0 \subseteq {\cal T}_0 \cup
{\cal K}$ with signature $\Pi = \Pi_0 \cup (\Sigma_P \cup \Sigma)$\\
& ~~~where $\Sigma_P$ is a set of parameters and ${\cal K}$ is a set of
flat and linear clauses; \\
& \!\!\!\!$G$, a finite set of flat and linear ground clauses in the signature $\Pi^C$; \\
& \!\!\!\!$T$, a finite set of flat ground $\Pi^C$-terms  s.t.\ ${\sf
  est}({\cal K}, G) \subseteq T$ and ${\cal K}[T]$ is ground. \\

{\bf Output:} & Universal  $\Pi_0 \cup \Sigma_P$-formula 
$\forall y_1 \dots y_n \Gamma_T(y_1,
\dots, y_n)$. \\
\hline 
\end{tabular}

\begin{description}
\vspace{-1mm}
\item[Step 1] Compute the set of $\Pi_0^C$ clauses $\K_0 {\cup} G_0 {\cup}
  {\sf Con}_0$  from $\K[T] \cup G$  using the purification step
  described in  Thm.~\ref{lemma-rel-transl} (with set of extension
  symbols $\Sigma_1 = \Sigma_P \cup \Sigma$). 


\item[Step 2]  $G_1 := {\mathcal K}_0 \cup G_0\cup {\sf Con}_0$. 
Among the constants in $G_1$, identify 
\begin{enumerate}
\item[(i)] the constants
${\overline c_f}$, $f {\in} \Sigma_P$, where $c_f {=} f {\in}
\Sigma_P$ is a constant
parameter or $c_f$ is 
introduced by a definition $c_f {:=} f(c_1,\dots,c_k)$ in the hierarchical
reasoning method, 
\item[(ii)] all constants  ${\overline c_p}$ 
occurring as arguments of functions in $\Sigma_P$ in such definitions. 
\end{enumerate}
Let ${\overline  c}$ be the remaining constants.\\
Replace the constants in ${\overline  c}$
with existentially quantified variables ${\overline x}$ in $G_1$,
i.e.\ 
replace $G_1({\overline c_p}, {\overline c_f}, {\overline c})$ 
with $G_1({\overline c_p}, {\overline c_f}, {\overline x})$, and
consider the formula
$\exists {\overline x} G_1({\overline c_p},{\overline c_f}, {\overline x})$.


\item[Step 3] Compute a quantifier-free formula 
$\Gamma_1({\overline c_p}, {\overline c_f})$  equivalent to 
$\exists {\overline x} G_1({\overline c_p}, {\overline c_f},{\overline
  x})$ w.r.t.\ $\T_0$ using a  method for quantifier elimination in 
${\mathcal T}_0$.  


\item[Step 4] Let $\Gamma_2({\overline c_p})$ be the formula 
obtained by replacing back in $\Gamma_1({\overline c_p}, {\overline c_f})$ 
the constants $c_f$ introduced by definitions $c_f := f(c_1, \dots,
c_k)$ with the terms $f(c_1, \dots,c_k)$. 

Replace ${\overline c_p}$ with existentially quantified variables ${\overline y}$. 


\item[Step 5] Let $\forall {\overline y} \Gamma_T({\overline y})$ be
  $\forall {\overline y} \neg \Gamma_2({\overline y})$. 
\vspace{-1mm}
\end{description}
\label{alg-symb-elim}
\end{algorithm}

\noindent 
We identify situations in which we can
generate, for every set of flat ground clauses $G$,   
a universal formula $\Gamma$ representing a family of
constraints on the parameters in $\Sigma_P$, such that $G$ is 
unsatisfiable w.r.t.\ ${\cal T}_0 \cup {\cal K} \cup {\Gamma}$.
 A possibility of doing this in a hierarchical way, by reducing 
the problem to quantifier elimination in the theory ${\cal T}_0$
is described in Algorithm~\ref{alg-symb-elim}.

\begin{thm}[\cite{sofronie-lmcs-2018,PeuterSofronie2019}] 
Assume that ${\mathcal T}_0$ allows quantifier elimination.  
Let ${\mathcal T}_0 \subseteq {\mathcal T}_0 \cup \K$ be an extension
of the theory ${\mathcal T}_0$ with additional function symbols in a
set $\Sigma_1 = \Sigma_P \cup \Sigma$ satisfying a set $\K$ of flat
and linear\footnote{A clause is flat if the arguments of function
  symbols are variables; it is linear if whenever a variable is
  a proper subterm in 
  different terms, the terms are equal.} clauses. 
Assume that $\K = \K_P \cup
\K_1$ such that $\K_P$ contains only function symbols in $\Sigma_0 \cup
\Sigma_P$ and $\K_1$ is a set of
   $\Pi$-clauses. Let $G$ be a set of flat and linear\footnote{A ground
  clause is flat if the arguments of function symbols are constants;
  it is linear if whenever a constant is a proper subterm in different terms,  the
  terms are equal.} ground $\Pi^C$-clauses such that parametric
constants do not occur below symbols in $\Sigma_1$ and $T$ a set of
flat $\Pi^C$-terms satisfying the conditions in Algorithm~1, and let  $\forall {\overline y}
\Gamma_T({\overline y})$ be the formula obtained applying Algorithm~\ref{alg-symb-elim}.
\begin{itemize}
\item[(1)] For every
$\Pi^C$-structure ${\mathcal A}$
which is a model of ${\mathcal T}_0 \cup {\mathcal K}$, 
if ${\mathcal A} \models \forall {\overline y}
\Gamma_T({\overline y})$ then ${\mathcal A} \models \neg G$, 
i.e.\  ${\mathcal T}_0 \cup {\mathcal K} \cup \forall {\overline y}
\Gamma_T({\overline y}) \cup 
  G$ is unsatisfiable. 
\item[(2)] 
Assume  that
${\cal T}_0 \subseteq {\cal T}_0 \cup {\cal K}_P \subseteq {\cal T}_0
\cup {\cal K}_P \cup \K_1$ 
is a chain of theory extensions both satisfying condition $({\sf
  Comp})$ and having the property that all variables occur below an
extension function, and such that $\K$ is flat and linear. 
Let 
$\forall {\overline y} \Gamma_1({\overline  y})$ be the formula obtained by 
applying Algorithm 1 to $\T_0 \cup \K_1$, $G$ and $T := {\sf
  est}(\K, G)$.  
Then the formula 
$\K_P \wedge 
\forall {\overline y} \Gamma_1({\overline y})$ 
has the property 
that for every universal formula $\Gamma$ containing only parameters
in $\Sigma_P$ with 
  ${\cal T}_0 \cup (\K_P \cup
  \Gamma)  \cup G \models \perp$, we have 
$\K_P \wedge \Gamma  \models  \K_P \wedge \forall {\overline y} \Gamma_1({\overline y})$. 
\end{itemize}
\vspace{-3mm}
\label{symb-elim-simplif}
\label{inv-trans-qe}
\end{thm}
A similar result can be established for $\Psi$-locality and 
for chains of local theory extensions, cf.\ also 
\cite{sofronie-lmcs-2018,PeuterSofronie2019,peuter-thesis-2024}. 

\begin{ex} Consider the extension of the theory of real numbers
 ${\mathbb R}$ with additional function symbols $L, L'$ satisfying axioms ${\cal K}$: 

\smallskip
${\cal K} = \{ \forall x (L(x) + m(x)*t \leq L'(x)), \quad \forall x
(L'(x) \leq L(x) + M(x)*t) \}.$

\smallskip
\noindent By the results in Section~\ref{examples} the theory extension satisfies condition ${\sf
  Comp}$.

Let $G := \{ t > 0, L(c) \leq l_{\sf max},  L'(c) > l_{\sf max} \}$.  
We use Algorithm~\ref{alg-symb-elim} with set of parameters $\Sigma_P
= \{ m, M \}$ to determine the weakest universal
condition $\Gamma$ on these parameters under which ${\mathbb R} \cup
{\cal K} \cup \Gamma \cup G  \models \perp$.

\begin{description}
\item[Step 1:] We instantiate all universally quantified variables in
  ${\cal K}$
with $c$. After replacing $L(c)$ with $d$, $L'(c)$ with $d'$, $m(c)$
with $e$ and $M(c)$ with $e'$ we obtain: ~~
$
(d + e*t \leq d') \wedge (d' \leq d + e'*t) \wedge t > 0 \wedge (d
\leq l_{\sf max}) \wedge (d' >  l_{\sf max}).
$

 \item[Step 2:] We distinguish the constants: (i) $e,
   e'$ introduced for terms starting with the parameter $m,
   M$, (ii) $c$ argument of parameters, and (iii) $t, d, d'$ which are
   regarded as existentially quantified variables.
   We consider the formula: 

$\exists t \exists d \exists d' ((d + e*t \leq d') \wedge (d' \leq d + e'*t) \wedge t > 0 \wedge (d
\leq l_{\sf max}) \wedge (d' >  l_{\sf max})).$

\item[Step 3:] We use quantifier elimination in ${\mathbb R}$  and obtain: $e \leq e' \wedge e' > 0$.

\item[Step 4:] 
We replace $e, e'$ back with $m(c)$ resp.\ $M(c)$ and regard $c$ as  existentially
quantified variable and obtain: $\exists c \Gamma_2(c) := \exists c (m(c) \leq M(c)
\wedge M(c) > 0)$.

\item[Step 5:] 
The negation is $\forall x (m(x) > M(x) \vee M(x) \leq 0)$. This is
the weakest universal additional condition under which $G$ does not hold. 

If ${\cal K}$ contains also $\forall x (m(x) \leq M(x))$, we can use this property to
simplify the formula computed in Step 4; so Step 5 would yield $\forall x (M(x) \leq 0)$. 
\end{description}
\end{ex}

\subsection{Tools}

\noindent {\bf H-PILoT.} The method for
hierarchical reasoning in local theory extensions described before
was implemented in the system \mbox{H-PILoT} \cite{hpilot}.
Standard SMT solvers such as CVC4, CVC5 or Z3 or specialized provers 
such as Redlog \cite{redlog} can be used for testing the satisfiability of the
formulae obtained after
the reduction to a satisfiability test w.r.t.\ the base theory.
The advantage in comparison with provers using heuristics for
instantiation directly is that knowing
the instances needed for a complete instantiation allows us to correctly detect
satisfiability (and generate models) in situations in which other SMT
provers return ``unknown''. Another advantage is that this complete 
instantiation can be further used for symbol elimination.

\medskip
\noindent {\bf SEH-PILoT.} 
SEH-PILoT (Symbol elimination with H-PILoT) is a tool that combines 
hierarchical reduction for local theory extensions with symbol 
elimination. This allows to automate the generation of constraints by 
implementing Algorithm 1, which can be further used for invariant strengthening 
(cf. e.g.\ \cite{PeuterSofronie2019}). 

SEH-PILoT is invoked with a YAML file that specifies tasks and
all options. A task is a description of a problem consisting, among other things
of a mode (satisfiability checking, constraint generation or invariant
strengthening); the base theory; 
 a list of parameters (or a list of symbols to be
eliminated);  task specific options such as a list of assumptions
which can be used for simplification; the formalization of the actual
problem in the syntax of H-PILoT. 
For the hierarchical reduction (Step 1 of Algorithm 1) SEH-PILoT utilizes H-PILoT. 
The result of H-PILoT is processed according to the task and to 
selected options for a solver to perform the symbol elimination. For 
tasks to generate constraints or strengthen invariants SEH-PILoT is 
limited at the moment to Redlog to perform the symbol elimination. The 
base theories are currently limited to the theory of real closed fields 
and the theory of Presburger arithmetic. SEH-PILoT is developed at the 
present to transform the obtained result of H-PILoT into SMT-LIB 
(version 2.7) to utilize a variety of additional state of the art solvers.

For each task, SEH-PILoT forms an appropriate invocation of 
H-PILoT according to the specification in the YAML file. Then it 
processes the output of H-PILoT (Step 2 of Algorithm 1) 
to form an appropriate file for the invocation of Redlog for
quantifier elimination (Step 3). Depending on the task 
specific options this file will be extended with additional Redlog 
commands. An example is the simplification of formulas using Redlog's 
interface to the external QEPCAD-based simplifier SLFQ or with a list of 
assumptions. After the invocation of Redlog on the generated file, the 
output is processed by SEH-PILoT to extract the required results
(Steps 4 and 5): 
The extracted formula representing the constraints or invariants is 
translated from the syntax of Redlog back to the syntax of H-PILoT, 
and symbols H-PILoT has introduced during hierarchical reduction are 
replaced back, such that the obtained formula does not contain new 
symbols. Depending on the chosen mode this is then either the final 
result of the task (a constraint) or the input for the next 
iteration (invariant strengthening). SEH-PILoT can generate, upon
request, statistics for all subtasks and steps of the process. 
The statistics indicate the time needed for the subtasks as well as
the number of atoms in the generated constraints before and after 
simplification with the external QEPCAD-based simplifier SLFQ.

\section{Parametric Linear Hybrid Automata}
\label{lha}

\label{Sec:HybridAutomata}

In this paper we present methods for the analysis of (families of) parametric linear
hybrid automata. We start in this section with a definition of hybrid automata and of
linear hybrid automata as given in \cite{Henzinger} and the
verification problems we consider. In Section~\ref{Sec:FamiliesLHA}  then we present
the way we defined systems of similar hybrid automata in
\cite{DammHS15}
and the related verification problems. 

Hybrid automata were introduced in \cite{Henzinger} to describe
systems with discrete control, such that in every control mode certain variables can
evolve continuously in time according to precisely specified rules.

\begin{defi}[Hybrid automaton \cite{Henzinger}]
A hybrid automaton is a tuple 

\smallskip 
$S = (X, Q, {\sf flow}, {\sf Inv}, {\sf Init}, E, {\sf guard}, {\sf jump})$

\smallskip
\noindent consisting of: 
\begin{itemize}
\vspace{-2mm}
\item[(1)] 
A finite set $X = \{ x_1, \dots, x_n \}$ of real valued variables
(whose values can change over time, and which are therefore regarded as functions
$x_i : {\mathbb R} \rightarrow {\mathbb R}$) and 
a finite set $Q$ of control modes; 
\item[(2)] A family $\{ {\sf flow}_q \mid q \in Q \}$ of 
predicates over the variables in
$X \cup {\dot X}$ (where ${\dot X} =\{ {\dot x_1}, \dots, {\dot x_n} \}$, 
where ${\dot x_i}$ is the derivative of $x_i$)  
specifying the 
continuous dynamics in each control mode\footnote{This means that we assume that the functions
$x_i : {\mathbb R} \rightarrow {\mathbb R}$ are differentiable during flows.}; 
a family $\{ {\sf Inv}_q \mid q \in Q \}$ of 
predicates over the variables in $X$ defining the  
invariant conditions for each control mode; 
and a family $\{ {\sf Init}_q \mid q \in Q \}$ of 
predicates over the variables in $X$, defining the initial 
states for each control mode. 
\item[(3)] A finite multiset $E$ 
with elements in $Q {\times} Q$ (the control switches). 
Every $(q, q') \in E$ is a 
directed edge between $q$ (source mode) and $q'$ (target mode); 
a family of guards 
$\{ {\sf guard}_e \mid e \in E \}$ (predicates over $X$); and a family of 
jump conditions $\{ {\sf jump}_e \mid e \in E \}$ 
(predicates over $X \cup X'$, 
where $X' =\{ x'_1, \dots, x'_n \}$ is a copy of $X$ consisting of ``primed'' 
variables). 
\end{itemize}
\end{defi}
A {\em state} of $S$ is a pair $(q, a)$ consisting
of a control mode $q \in Q$ and a vector $a = (a_1, \dots, a_n)$
that represents a value $a_i \in {\mathbb R}$ for each variable $x_i \in X$.
A state $(q, a)$ is {\em admissible} if ${\sf Inv}_q$ 
is true when each $x_i$ is replaced by $a_i$.
There are two types of {\em state change}: 
(i) A {\em jump} is an instantaneous transition 
that changes the control location and the values of variables in $X$ according
to the jump conditions;   
(ii) In a {\em flow}, the state can change due to the evolution in a 
given control mode over an interval of time: the values of the
variables in $X$  
change continuously according to the flow rules of the current
control location; all intermediate states are admissible.
A {\em run} of $S$ is a finite
sequence $s_0 s_1 \dots s_k$ of  admissible states such that
(i) the first state $s_0$ is an initial state of $S$ (the values of 
the variables satisfy ${\sf Init}_q$ for some $q \in Q$),
(ii) each pair $(s_j, s_{j+1})$ is either a jump of $S$ or the endpoints of a 
flow of $S$.

\noindent {\em Notation.} In what follows we use the following notation.
If $x_1, \dots, x_n \in X$ we denote 
the sequence $x_1, \dots, x_n$ with ${\overline x}$, the sequence 
$\dot{x}_1, \dots, \dot{x}_n$ with $\overline{\dot{x}}$, and  
the sequence of values $x_1(t), \dots, x_n(t)$ of these variables 
at a time $t$ with ${\overline x}(t)$.

\medskip

\noindent In \cite{Henzinger} a class of hybrid automata was introduced in which the
flow conditions, the guards and the invariants have a special form. 

\begin{defi}
Let $X = \{ x_1,
\dots, x_n \}$ be a set of variables. An (atomic) linear predicate on the variables $x_1,
\dots, x_n$ is a linear strict or non-strict
inequality
of the form 
$a_1 x_1 + \dots a_n x_n \rhd a$, 
where $a_1, \dots, a_n,
a \in {\mathbb R}$ and $\rhd \in \{ \leq, < , \geq, > \}$. 
A convex linear predicate is a finite conjunction of linear
inequalities.
\end{defi} 

\begin{defi}[Linear hybrid automaton \cite{Henzinger}]
A hybrid automaton $S$  is
a linear hybrid automaton (LHA) if it satisfies the following
two requirements: 

\smallskip
\noindent {\em 1. Linearity:} For every control mode $q \in Q$, 
the flow condition ${\sf flow}_q$, the invariant condition ${\sf Inv}_q$, 
and the initial condition ${\sf Init}_q$ are convex linear 
predicates.
For every control switch $e = (q,q') \in E$, the jump condition
${\sf jump}_e$ and the guard ${\sf guard}_e$ are convex linear 
predicates.
In addition, 
as in \cite{damm-ihlemann-sofronie-hscc11,DammIS11}, 
we assume that the 
flow conditions ${\sf flow}_q$ are conjunctions of {\em non-strict} 
linear inequalities. 

\smallskip
\noindent {\em 2. Flow independence:} For every control mode $q \in Q$, 
the flow condition ${\sf flow}_q$ is a predicate over the variables
in ${\dot X}$ only (and does not contain any variables from $X$).
This requirement ensures that the possible flows
are independent from the values of the variables, and only depend
on the control mode.
\end{defi} 

\medskip
\noindent 
\begin{defi}[Parametric linear hybrid automaton \cite{DammHS15}]
A parametric hybrid automaton (PLHA) is a
linear hybrid automaton for which a set $\Sigma_P = P_c \cup P_f$ of 
parameters is specified (consisting of parametric constants $P_c$ and 
parametric functions $P_f$)
with the difference that for every control mode $q \in Q$ and every mode 
switch $e$: 
\begin{itemize}
\item[(1)] the linear constraints in the invariant conditions ${\sf Inv}_q$, 
initial conditions ${\sf Init}_q$, and 
guard conditions ${\sf guard}_e$ are of the form: 
$g \leq \sum_{i = 1}^n a_i x_i \leq f$,

\item[(2)] the inequalities in the 
flow conditions ${\sf flow}_q$ are of the form:  
$\sum_{i = 1}^n b_i {\dot x}_i \leq b$,

\item[(3)] the linear constraints in ${\sf jump}_e$ are of the form 
$\sum_{i = 1}^n b_i x_i + c_i x'_i \leq d$,
\end{itemize}
\noindent (possibly relative to an interval $I$) 
where the coefficients $a_i, b_i, c_i$ and the bounds $b, d$ 
are either numerical constants or parametric constants in $P_c$; 
and $g$ and $f$ are (i) constants or parametric constants in $P_c$, or 
(ii) parametric functions in $P_f$ satisfying 
the convexity (for $g$) resp. concavity condition (for $f$), or 
terms with one free variable $t$ such that the associated functions
have these convexity/concavity properties 
and $\forall t (g(t) \leq f(t))$. 
The flow independence conditions hold as 
in the case of linear hybrid automata.
\end{defi}

\subsection{Verification}  
We consider 
the problem of checking whether a quan\-ti\-fier-free formula $\Phi$ 
in real arithmetic 
over the variables $X$ is an inductive invariant in a 
hybrid automaton $S$, i.e.: 
\begin{itemize}
\item[(1)] $\Phi$ holds in the initial states of mode $q$ for all $q \in Q$; 
\item[(2)] $\Phi$ is invariant under jumps and flows:
\begin{itemize}
\item For every flow in a mode $q$, 
the continuous variables satisfy 
$\Phi$ both during and at the end of the flow. 
\item  For every jump, 
 if the values of the continuous variables satisfy 
$\Phi$ before the jump, they satisfy $\Phi$ after the jump.
\end{itemize}
\end{itemize}

\begin{thm}[\cite{DammIS11}]
Let $S$ be a LHA with real-valued variables $X$ and $\Phi$ a property
expressible as a convex linear predicate over $X$. The following are equivalent: 
\begin{itemize}
\vspace{-1mm}
\item[(1)] $\Phi$ is an inductive invariant of the hybrid automaton;  

\item[(2)] 
For every $q \in Q$ and $e = (q,q') \in E$, the following formulae are 
unsatisfiable:  
\end{itemize}

\noindent $\begin{array}{@{}ll@{}}
I_q  & {\sf Init}_q({\overline x})  \wedge \neg \Phi({\overline x}) \\[1ex] 

F_{\sf flow}(q) & \Phi({\overline x}(t_0)) \wedge {\sf Inv}_q({\overline x}(t_0)) \wedge 
{\underline {\sf flow}}_q(t_0, t) \wedge 
 {\sf Inv}_q({\overline x}(t)) \wedge 
\neg \Phi({\overline x}(t)) \wedge t \geq t_0 \\[1ex]

F_{\sf jump}(e)~~~~~ & 
\Phi({\overline x}(t)) \wedge {\sf jump}_e({\overline x}(t), {\overline x}'(0)) \wedge  
{\sf Inv}_{q'}({\overline x}'(0)) \wedge 
\neg \Phi({\overline x}'(0))
\end{array}$

\smallskip
\noindent where if ${\sf flow}_q  =  
\bigwedge_{j = 1}^{n_q} (\sum_{i = 1}^n c^q_{ij} {\dot x}_i \leq_j
c_j^q)$ then: 

\smallskip
\noindent ${\underline {\sf flow}}_q(t, t') = 
\bigwedge_{j = 1}^{n_q}  
(\sum_{i = 1}^n c^q_{ij} (x_i' - x_i) \leq_j c_j^q (t' - t))$, where
$x'_i = x_i(t'), x_i = x_i(t)$.
\vspace{-2mm}
\label{transl-invar-par}
\end{thm}
As a consequence of Theorem~\ref{transl-invar-par}, we can determine 
the complexity of verification of LHA, and the complexity of constraint
generation for PLHA.  
\begin{cor}
Let $S$ be a (P)LHA with real-valued variables $X$ and $\Phi$ a property
expressible as a convex linear predicate over $X$. 
\begin{itemize}
\item[(1)] {\bf Verification of LHA.} Assume that all coefficients used in the convex linear
  predicates in the description of $S$ are concrete constants.
Then the problem of checking whether $\Phi$ is an
inductive invariant  is decidable in PTIME \footnote{By
  Theorem~\ref{transl-invar-par}, the problem can be reduced to checking the satisfiability of a
family of conjunctions of linear inequalities in linear real
arithmetic which is linear in the size of the description of $S$ and
of $\Phi$; the satisfiability over ${\mathbb R}$ of conjunctions of linear inequalities can be checked in PTIME
\cite{khakian}.}.
\item[(2)] {\bf Verification/Constraint generation for PLHA.} Assume that some of the coefficients used in the convex linear
  predicates in the definition of $S$ are parametric constants,
  and additional constraints on these parameters are specified.\footnote{By
  Theorem~\ref{transl-invar-par}, the problem can be reduced to checking the satisfiability of a
family of conjunctions of non-linear atoms which is linear in the size of the description of $S$ and
of $\Phi$. Checking satisfiability of non-linear inequalities and
quantifier elimination can be done in exponential time.}
The problem of checking whether $\Phi$ is an inductive
  invariant is decidable in exponential time. 
Determining constraints on the parameters under which
  $\Phi$ is guaranteed to be an inductive invariant can be done in
  exponential time by 
  quantifier elimination in the theory of real closed
  fields. 
\end{itemize}
\end{cor}

\subsection{Example: Verification and constraint generation}

We now illustrate the ideas presented before on variants of Example~\ref{ex1}.

\begin{ex}
{\em Consider a water tank modelled as described in
Example~\ref{ex1} using a hybrid system 
with variable $L$ (water level) and two modes $s_1, s_2$ (state invariants 
$L \geq L_{\sf a}$ and $L < L_{\sf a}$, where $L_{\sf a}$ is an alarm
level). 
The water level, as well as the inflow and outflow are modeled using
unary functions $L$, ${\sf infl}, {\sf outfl}$, where
$L(t)$, ${\sf infl}(t)$ and ${\sf outfl}(t)$ are the water level,  the
inflow and  outflow at time $t$, respectively. 
We here assume that the inflow and outflow rates are constant 
and equal to ${\sf in}$, resp.\ ${\sf out}$ (i.e.\ the derivative of
${\sf infl}$ is equal to ${\sf in}$ at every point in time $t$ and the derivative of
${\sf outfl}$ is equal to ${\sf out}$ at every point in time $t$). 
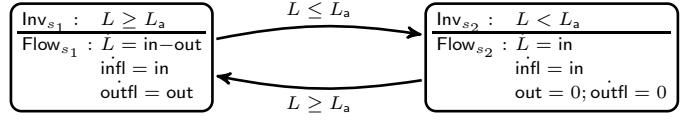
\begin{center}
{\scriptsize 
\begin{tikzpicture}
\node [square state] (1) {$\begin{array}{ll}
{\sf Inv}_{s_1}: \quad & L \geq L_{\sf a}\\
\hline 
{\sf Flow}_{s_1}: & \dot{L} = {\sf in} {-} {\sf out}\\
& \dot{{\sf infl}} = {\sf in}\\
& \dot{{\sf outfl}} = {\sf out}
\end{array}$}; 
\node [square state] (2) [right = of 1]{$\begin{array}{ll}
{\sf Inv}_{s_2}: \quad  & L <
L_{\sf a}\\
\hline 
{\sf Flow}_{s_2}: & \dot{L} = {\sf in}\\
& \dot{{\sf infl}} = {\sf in}\\
& {\sf out} = 0; \dot{{\sf
    outfl}} = 0
\end{array}$};
 
\path 
(1) edge [bend left=10,above] node  {$L \leq L_{\sf a}$} (2)
(2) edge [bend left=10,below] node  {$L \geq L_{\sf a}$} (1);  
\end{tikzpicture}
}
\end{center}

\noindent Assume that  ${\sf Init}_{s_1} = (L {=} L_1) \wedge (L
{\geq} L_{\sf a})$ and ${\sf Init}_{s_2}
= (L {=} L_2) \wedge (L {<} L_{\sf a})$. 

\noindent Consider the safety condition $\Phi = L \leq L_{\sf o}$ stating that
the water level always remains below an overflow level, $L_{\sf o}$. 
To prove that $\Phi$ is an inductive invariant, by
Theorem~\ref{transl-invar-par}, we need to prove that
the following formulae are unsatisfiable: 
\begin{itemize}
\item[(1)] ${\sf Init}_{s_1} \wedge \neg \Phi$ \\
${\sf Init}_{s_2} \wedge \neg \Phi$
\item[(2)] $\Phi(L(t_0)) \wedge {\sf Inv}_{s_1}(L(t_0)) \wedge
  {\underline {\sf flow}}_{s_1}(t_0, t) \wedge {\sf Inv}_{s_1}(L(t))
  \wedge \neg \Phi(L(t)) \wedge t \geq t_0$ \\
$\Phi(L(t_0)) \wedge {\sf Inv}_{s_2}(L(t_0)) \wedge
  {\underline {\sf flow}}_{s_2}(t_0, t) \wedge {\sf Inv}_{s_2}(L(t))
  \wedge \neg \Phi(L(t)) \wedge t \geq t_0$ 
\item[(3)] $\Phi(L) \wedge {\sf guard}_{s,s'}(L) \wedge L' = L \wedge
  \neg \Phi(L')$, where $s, s' \in \{ s_1, s_2 \}$ and $s \neq s'$,
\end{itemize}
\begin{tabular}{@{}l@{}l}
where~ & ${\underline {\sf flow}}_{s_1}(t_0, t) :=  L(t) - L(t_0) = ({\sf
  in} - {\sf out})*(t - t_0)$ and  \\
& ${\underline {\sf flow}}_{s_2}(t_0, t) :=  L(t) - L(t_0) = {\sf
  in}*(t - t_0)$. 
\end{tabular}

\medskip
\noindent (1) {\bf Checking whether $\Phi$ holds in the initial
  states.} To check that the formula $L \leq L_{\sf o}$ is an
inductive invariant, we first check
whether it holds in the initial states, i.e.\ check whether 
$L {=} L_1 \wedge L {\geq} L_{\sf a} \wedge L {>}  L_{\sf o}$ and 
$L {=} L_2 \wedge L {<} L_{\sf a} \wedge L {>}  L_{\sf o}$
are unsatisfiable. Without additional assumptions about $L_1, L_2,
L_{\sf a}$ and $L_{\sf o}$ these formulae are satisfiable, so if we consider
the constants $L_1, L_2, L_{\sf a}$ and $L_{\sf o}$ to be parameters, 
we can derive conditions on these parameters under which the 
formulae are guaranteed to be unsatisfiable i.e.\ $\Phi$ is guaranteed
to hold in the initial states. 

\noindent The conditions can be derived by eliminating the variable $L$, i.e.\ 
by computing:

\smallskip
$\exists L ((L = L_1) \wedge  (L
\geq L_{\sf a}) \wedge (L > L_{\sf o})) \equiv (L_1 > L_{\sf o}) \wedge
(L_1 \geq L_{\sf a})$

$\exists L ((L = L_2) \wedge  (L
< L_{\sf a}) \wedge (L > L_{\sf o})) \equiv (L_2 > L_{\sf o}) \wedge
(L_2 < L_{\sf a})$

\smallskip
\noindent and then negating the result. We obtain the conditions: 

\smallskip
$(L_1 \leq L_{\sf o}) \vee
(L_1 < L_{\sf a})$ and $(L_2 \leq L_{\sf o}) \vee
(L_2 \geq L_{\sf a})$. 

\smallskip
\noindent If we assume in addition that the initial states satisfy the
invariants of the respective modes, i.e.\ that $L_1 \geq L_{\sf a}$ and
$L_2 < L_{\sf a}$, then the condition on $L_1, L_2$ under which 
in the initial states the formula $\Phi$ holds is $(L_1 \leq L_{\sf o}) \wedge
(L_2 \leq L_{\sf o})$. 

\medskip
\noindent (2) {\bf Checking invariance under flows.} The formula $L \leq L_{\sf o}$ is invariant under flows iff the formulae in (2)
are unsatisfiable w.r.t.\ the extension of the theory ${\cal T}_S$ of
real numbers with a function symbol $L$ satisfying the axiom $\forall t\, L(t)
\geq 0$:  

\begin{itemize}
\item[(i)] $ L(t_0) \leq L_{\sf o} ~\wedge~ t_0 < t
 ~ \wedge~ L(t_0) \geq L_{\sf a}  ~\wedge~ L(t) \geq L_{\sf a}
  ~\wedge~ L(t) > L_{\sf o} ~\wedge \\
 L(t) - L(t_0) = ({\sf in} {-}
  {\sf out}) {*}(t - t_0) $,
\item[(ii)]  $L(t_0) \leq L_{\sf o} ~\wedge~ t_0 < t ~\wedge~ L(t_0) <
  L_{\sf a}  ~\wedge~ L(t) < L_{\sf a} ~\wedge~ L(t) > L_{\sf o}  ~\wedge \\
L(t) - L(t_0) = {\sf in}{*}(t - t_0) $.
\end{itemize}
Since the extension of the theory ${\mathbb R}$ of real numbers with
the function symbol $L$ satisfying condition $\forall t\, L(t) \geq 0$
is local, we can use the method
for hierarchical reasoning described in Theorem~\ref{lemma-rel-transl} for checking
the satisfiability of these formulae. We proceed as follows: 
We introduce new constants $l, lp$ with their definitions: ${\sf
  Def} = \{ l = L(t_0), lp = L(t) \}$. 
Then the formulae in (i) and (ii) are
unsatisfiable iff the formulae (i') and (ii') below are unsatisfiable:

\begin{itemize}
\item[(i')] $ l \leq L_{\sf o} \wedge t_0 < t
  \wedge l \geq L_{\sf a} \wedge 
 lp - l = ({\sf in} {-}
  {\sf out}) {*}(t - t_0) \wedge lp \geq L_{\sf a}
  \wedge lp > L_{\sf o}$,
\item[(ii')]  $l \leq L_{\sf o} \wedge t_0 < t \wedge l <
  L_{\sf a} \wedge  
lp - l = {\sf in}{*}(t - t_0) \wedge lp < L_{\sf a} \wedge lp > L_{\sf o}$.
\end{itemize}
Note that $t_0 < t \models {\sf Con} = (t_0 = t
\rightarrow l = lp)$, so the instances of the congruence axioms 
are not needed in this case.
The satisfiability of (i) and (ii) can be checked with H-PiLoT. 
The satisfiability of (i') and
(i'') can be checked with a prover for the theory of real
numbers. 

\noindent Since the formulae are satisfiable, $\Phi$ is
not invariant under flows without additional assumptions on 
$L_{\sf o}, L_{\sf a}, {\sf in}, {\sf out}$. 

\smallskip
\noindent We can use Algorithm~1 to determine the weakest conditions on the parameters 
$\Sigma_P = \{ L_{\sf o}, L_{\sf a}, {\sf in}, {\sf out} \}$ 
which guarantee the invariance of $\Phi$
under flows as follows:

\medskip
\noindent (2.i) {\bf Invariance under flows in mode $s_1$:} 
\begin{description}
\item[Step 1:] We start with the formula in (i') obtained after
  instantiation and purification.

\item[Step 2:] Among the constants in this formula, we identify 
the parameters in $\Sigma_P$ (which do not have to be eliminated) 
$ \{ L_{\sf o}, L_{\sf a}, {\sf in}, {\sf out} \}$,
and the constants $t_0, t, l, lp$, which have
to be eliminated.  

\item[Step 3:] To eliminate $t_0, t, l, lp$ note that: 
\end{description}

\noindent {\small $\exists t_0, t \exists l, lp ( l \leq L_{\sf o} \wedge t_0 < t
  \wedge l \geq L_{\sf a} \wedge 
 lp {-} l = ({\sf in} {-}
  {\sf out}) {*}(t {-} t_0) \wedge lp \geq L_{\sf a}
  \wedge lp > L_{\sf o})$} 

\noindent $\equiv  \exists t_0, t ( L_{\sf a} \leq L_{\sf o} \wedge t_0 < t \wedge L_{\sf a} \leq
L_{\sf o} \wedge$ \\
$~~~~~~~~~~~~L_{\sf a} - ({\sf in} - {\sf out})(t - t_0) \leq L_{\sf o}
\wedge L_{\sf o} - ({\sf in} - {\sf out})(t - t_0) < L_{\sf o})$ 

\noindent $\equiv (L_{\sf a} \leq L_{\sf o} \wedge {\sf in} - {\sf
  out} > 0)$

\begin{description}
\item[Step 4:] We negate the formula obtained in Step 3 and obtain: 

$L_{\sf o} < L_a \vee {\sf in} - {\sf out} \leq 0$.

If we assume that $L_{\sf a} \leq L_{\sf o}$, then the condition above
can be simplified to ${\sf in} - {\sf out} \leq 0$.
\end{description}

\noindent The tests with SEH-PILoT in which positivity conditions for 
$L_{\sf a},$ $L_{\sf o},$ ${\sf in}, {\sf out}$ are
included as assumptions (and can be used for the simplification of
formulae) can
be found below (we used {\sf i} instead of {\sf in} and {\sf o}
instead of {\sf out} because of syntactic restrictions in Redlog): 

\smallskip
{\scriptsize 
\begin{lstlisting}[frame=single]
tasks:
  water-tanks-sat-constraint_slfq:
        mode: GENERATE_CONSTRAINTS
        solver: REDLOG
        options:
            parameter: [i,o,la,lo]
            assumptions: [t0 < t1,0 < i,0 <= o,0 < la,0 < lo]
            slfq_query: true
        specification_type: HPILOT
        specification_theory: REAL_CLOSED_FIELDS
        specification: &spec_water-tanks-sat
            file: |
               Base_functions := {(-,2,0,real),(+,2,0,real),(*,2,0,real)}
               Extension_functions := {(l, 1, 1)}
               Relations := {(<,2),(<=,2),(>,2),(>=,2)}
               Constants := {(t0, real), (t1, real), (i, real), 
                             (o, real), (la, real), (lo, real)}
               Clauses := 
               (FORALL t). l(t) >= _0;
               Query :=
               t0 < t1; 
               l(t0) < lo;
               l(t0) >= la;
               l(t1) = l(t0) + ((i - o)*(t1 - t0));
               l(t1) >= la;
               % Negated safety condition:
               l(t1) > lo;
\end{lstlisting}

\

\begin{lstlisting}[frame=single]
Metadata:
    Date: '2025-04-10 17:15:38'
    Number of Tasks: 1
    Runtime Sum (s): 0.734
water-tanks-sat-constraint_slfq:
    Result: OR(la - lo >= _0, i - o <= _0)
    Runtime (s): 0.734
\end{lstlisting}
}

\smallskip
\noindent The test with SEH-PILoT with the additional assumption $L_{\sf a} < L_{\sf o}$ yields: 

\smallskip
{\scriptsize 
\begin{lstlisting}[frame=single]
Metadata:
    Date: '2025-04-10 17:06:28'
    Number of Tasks: 1
    Runtime Sum (s): 0.7711
water-tanks-sat-constraint_slfq:
    Result: i - o <= _0
    Runtime (s): 0.7711
\end{lstlisting}
}

\medskip
\noindent (2.ii) {\bf Invariance under flows in mode $s_2$, when
  $\Sigma_P = \{ L_{\sf o}, L_{\sf a}, {\sf in}, {\sf out} \}$:} 
\begin{description}
\item[Step 1:] We start with the formula in (ii') obtained after
  instantiation and purification.

\item[Step 2:] Among the constants in this formula, we identify 
the parameters in $\Sigma_P$ (which do not have to be eliminated) 
$ \{ L_{\sf o}, L_{\sf a}, {\sf in}, {\sf out} \}$,
and the constants $t_0, t, l, lp$, which have
to be eliminated.  

\item[Step 3:] A quantifier-free formula equivalent to 
\end{description}

\noindent $\exists t_0, t \,\exists l, lp \,(l \leq L_{\sf o} \wedge t_0 < t \wedge l <
  L_{\sf a} \wedge  
lp - l = {\sf in}{*}(t - t_0) \wedge lp < L_{\sf a} \wedge lp >
L_{\sf o})$

\noindent $\equiv \exists t_0, t (t_0 {<} t \wedge
{\sf in}*(t-t_0) > 0 \wedge L_{\sf o} - {\sf in}*(t-t_0) < L_{\sf a}
\wedge L_{\sf o} < L_{\sf a})$

\noindent $\equiv (L_{\sf o} < L_{\sf a} \wedge {\sf in} > 0)$

\begin{description}
\vspace{-2mm}
\item[Step 4:] We negate the condition obtained in Step 3 and obtain
  $L_{\sf a} \leq L_{\sf o} \vee {\sf in} \leq 0$, which is equivalent
  to $L_{\sf a} \leq L_{\sf o}$ under the additional assumption that
  ${\sf in} > 0$.
\end{description}
Alternatively, we might decide to allow also $t_0, t$ as parameters. 
Then in Step 3 we do not eliminate $t_0$ and $t$. The constraint 
obtained in Step 3 is: 

\smallskip
$t_0 < t ~\wedge~
{\sf in}*(t-t_0) > 0 ~\wedge~ L_{\sf o} - {\sf in}*(t-t_0) < L_{\sf a}
~\wedge~ L_{\sf o} < L_{\sf a}.$

\smallskip
\noindent The results obtained with SEH-PiLoT can be found below: 

\noindent Tests with $\Sigma_P = \{ L_{\sf a}, L_{\sf o}, {\sf in}, {\sf out}
\}$ and positivity conditions on the parameters.  

\smallskip
{\scriptsize 
\begin{lstlisting}[frame=single]
tasks:
  water-tanks-sat-constraint_slfq:
        mode: GENERATE_CONSTRAINTS
        solver: REDLOG
        options:
            parameter: [i,o,la,lo]
            assumptions: [t0 < t1, 0 < i, 0 <= o, 0 < la, 0 < lo]
            slfq_query: true
        specification_type: HPILOT
        specification_theory: REAL_CLOSED_FIELDS
        specification: &spec_water-tanks-sat
            file: |
               Base_functions := {(-,2,0,real),(+,2,0,real),(*,2,0,real)}
               Extension_functions := {(l, 1, 1)}
               Relations := {(<,2),(<=,2),(>,2),(>=,2)}
               Constants := {(t0, real), (t1, real), (i, real), 
                             (o, real), (la, real), (lo, real)}
               Clauses := 
               (FORALL t). l(t) >= _0;
               Query :=
               t0 < t1; 
               l(t0) <= lo;
               l(t0) < la;
               l(t1) = l(t0) + (i*(t1 - t0));
               l(t1) < la;
               % Negated safety condition:
               l(t1) > lo;
\end{lstlisting}

\

\begin{lstlisting}[frame=single]
Metadata:
    Date: '2025-04-11 13:23:33'
    Number of Tasks: 1
    Runtime Sum (s): 0.6996
water-tanks-sat-constraint_slfq:
    Result: la - lo <= _0
    Runtime (s): 0.6996
\end{lstlisting}
}

\smallskip
\noindent Tests with $\Sigma_P = \{ L_{\sf a}, L_{\sf o}, {\sf in}, {\sf out}
\}$, positivity conditions on the parameters, and the condition
$L_{\sf a} \leq L_{\sf o}$:

\smallskip
{\scriptsize 
\begin{lstlisting}[frame=single]
Metadata:
    Date: '2025-04-11 13:21:40'
    Number of Tasks: 1
    Runtime Sum (s): 0.7289
water-tanks-sat-constraint_slfq:
    Result: 'true'
    Runtime (s): 0.7289
\end{lstlisting}
}

\smallskip
\noindent Thus, if $L_{\sf a} \leq L_{\sf o}$ formula (ii) is already
unsatisfiable. 

\ignore{\smallskip
\noindent Tests with $\Sigma_P = \{ L_{\sf a}, L_{\sf o}, {\sf in},
{\sf out}, t_0, t\}$, and only positivity conditions on the parameters:

\smallskip
{\scriptsize 
\begin{lstlisting}[frame=single]
asks:
  water-tanks-sat-constraint_slfq:
        mode: GENERATE_CONSTRAINTS
        solver: REDLOG
        options:
            parameter: [i,o,la,lo,t0,t1]
            assumptions: [t0 < t1, 0 < i, 0 <= o, 0 < la, 0 < lo]
            slfq_query: true
        specification_type: HPILOT
        specification_theory: REAL_CLOSED_FIELDS
        specification: &spec_water-tanks-sat
            file: |
                Base_functions := {(-,2,0,real), (+,2,0,real), (*,2,0,real)}
                Extension_functions := {(l, 1, 1)}
                Relations := {(<,2),(<=,2),(>,2),(>=,2)}
                Constants := {(t0, real), (t1, real), (i, real), 
                                       (o, real), (la, real), (lo, real)}
                Clauses := 
                (FORALL t). l(t) >= _0;
                Query :=
                t0 < t1; 
                l(t0) <= lo;
                l(t0) < la;
                l(t1) = l(t0) + (i*(t1 - t0));
                l(t1) < la;
                % Negated safety condition:
                l(t1) > lo;
\end{lstlisting}

\

\begin{lstlisting}[frame=single]
Metadata:
    Date: '2025-04-11 13:30:15'
    Number of Tasks: 1
    Runtime Sum (s): 0.7311
water-tanks-sat-constraint_slfq:
    Result: OR(la - lo <= _0, ((i * t0) - (i * t1)) + la <= _0)
    Runtime (s): 0.7311
\end{lstlisting}
}
}

\medskip
\noindent (3) {\bf Checking invariance under jumps.} Since in the mode changes $L$ is not updated, the formulae in (3)
above are unsatisfiable, so $\Phi$ is clearly
invariant under jumps. 

} 
\end{ex}

\section{Families of Similar Hybrid Automata}
\label{Sec:FamiliesLHA}

We present a possibility of describing families $\{ S(i) \mid i \in I
\}$ consisting of 
an unbounded number of similar (but not necessarily identical) 
hybrid automata proposed in
\cite{DammHS15}. 
To describe such families, we have to specify 
the properties of the component systems and their interaction.  

The systems $S(i)$ are hybrid automata; their interaction can be
described using a finite set of unary function symbols 
 which model the way the systems perceive  
other systems using 
 sensors in $P_S$, or by neighborhood connections 
(e.g. established by communication channels) in $P_N$. 
The structures modeling the topology of the system have the form 
$(I, \{ p : I \rightarrow I \}_{p \in P})$ where $P = P_S \cup P_N$.

\medskip
\noindent {\bf Component Systems.} 
We consider families of hybrid automata $\{ S(i) \mid i \in I \}$, with the same 
set of control modes $Q$ and the same 
mode switches $E \subseteq Q \times Q$, 
and whose real valued variables $X_{S(i)}$ are partitioned into a set 
$X(i) = \{ x(i) \mid x \in X \}$ of variables describing the
states of the system $S(i)$ and  a set $X_P(i) = \{ x_p(i) \mid x \in
X, p \in P \}$ describing the state of the neighbors $\{ p(i)  \mid p
\in P \}$  of $i$,  
where $X = \{ x_1, \dots, x_n \}$. 
We assume that all sets $X(i), i \in I$ are disjoint. 
Every component system $S(i)$ has the form: 
\newcommand{\variablesofautomaton}{X(i) \cup X_P(i)}
\newcommand{\derivativesofautomaton}{\{\variable[.]{i},
  \variable[.]{\pointer{i}} \mid \variable{}\in X, \pointer{} \in P\}}

\medskip
$ S(i) = (\variablesofautomaton, Q, \flow(i), \invariant(i), \initialStates(i), E, \guard(i), \jump(i))$ 

\medskip
\noindent 
where  for every $q \in Q$ and $e \in E$ 
${\sf flow}_q(i)$, ${\sf Inv}_q(i)$, ${\sf Init}_q(i)$,  ${\sf guard}_e(i)$,
 $ {\sf jump}_e(i)$
are conjunctions of formulae of the form ${\cal E} \vee {\cal C}$,
where ${\cal C}$ is a predicate over 
$X_{S(i)}$ (for  $\invariant(i), \initialStates(i), \guard(i)$),
or over $X_{S(i)} \cup \dot{X}_{S(i)}$ (for $\flow(i)$) resp. over 
$X_{S(i)} \cup X'_{S(i)}$ (for $\jump(i)$) and ${\cal E}$ is a disjunction 
of definedness conditions for the terms $p(i)$ occurring in ${\cal C}$ (for
instance, if for modeling the neighbors we use a theory of pointers as 
explained in Section~\ref{examples}, page~\pageref{remark-pointers}, ${\cal E}$ is a disjunction of 
equalities of the form $i = \nil$ and $p(i) = \nil$ if $x_p(i)$ occurs
in ${\cal C}$). 
For all $i \in I$ these formulae differ only in the variable index.
We consider two possibilities for $x_p(i)$:  
\begin{itemize}
\vspace{-1mm}
\item[(a)] $x_p(i)$ is at any moment the value of $x(p(i))$, the value 
  of variable $x$ for the system $S(p(i))$ and is controlled by suitable flow/jump
  conditions of $S(p(i))$; 
\item[(b)] $x_p(i)$ is the value of $x(p(i))$ which was sensed by the 
  sensor in the last measurement, and does not change between 
measurements. 
\label{a-b}
\vspace{-1mm}
\end{itemize}
We say that the system $S(i)$ is {\em linear} if 
\begin{itemize}
\item[(i)] $\flow(i)$ contains only variables in
$\dot{X}_{S(i)}$ and 
\item[(ii)] $\flow(i),$ $\invariant(i),$ $\initialStates(i),$
$\guard(i), \jump(i)$ are conjunctions of formulae ${\cal E} \vee
{\cal C}$,
as above, where ${\cal C}$ is a linear inequality (non-strict for flows)
and ${\cal E}$ is a disjunction of definedness conditions for the
terms $p(i)$ occurring in ${\cal C}$, as explained above. 
\end{itemize}
\noindent We consider systems of {\em parametric} LHA, in which
some coefficients or bounds in the linear inequalities 
are parameters in a set $\Sigma_{\sf Par}$.

\medskip
\noindent {\bf Topology.}
The topology of the family of systems and its updates was modeled in 
\cite{DammHS15} using an automaton 
$\environment$ with one mode, having 
as read-only-variables all  variables in $\{ x(i) \mid x \in X, i
\in I \}$ and as write variables  $\{ p(i) \mid p \in P, i\in I \}$, where $P
= P_S \cup P_N$.  
\ignore{The  initial states ${\sf Init}$ are described using ${\cal L}_{{\sf
    p}, {\sf n}}$-formulae.  
The jumps can represent updates of the sensor values $p(i), p 
\in P_S$ for a single system $S(i)$, 
but also synchronized 
global updates of the sensors $p \in P_S$ or neighborhood connections $p \in P_N$ for subsets 
of systems with a certain property (described by a 
formula): This can be useful when modeling 
systems of systems with an external controller (e.g.\ systems of car
platoons)  and entails 
a simultaneous update of an unbounded set of variables. 
Therefore, the description of the mode switches (topology updates) 
in $\environment$ is of a global nature.} 
The description of mode switches (topology updates) is of a global
nature; the update rules for $p \in P$, ${\sf Update}(p, p')$, are conjunctions of
implications: 
\begin{eqnarray}
\forall i (i \neq \nil \wedge \phi^p_k(i) \rightarrow F^p_k(p'(i), i)),
\quad \quad k \in \{ 1, \dots, m \} 
\label{updates-top}
\end{eqnarray} 
which describe how the values of the pointer $p$ change depending on a
set of mutually exclusive conditions $\{ \phi^p_1(i), \dots,\phi^p_m(i) \}$.
The variables 
$\{ x(i) \mid x {\in} X, i {\in} I \}$ can be used in the guards of ${\sf Update}(p, p')$, 
but cannot be updated by $\environment$. 
If $x_p(i)$ stores the value of $x(p(i))$ at the update of $p$ (case
(b) on page~\pageref{a-b}),  
then the update rules also change $x_p(i)$, so  $F^p_k(p'(i), i)$ 
must contain $x'_p(i) = x(p'(i))$ as a conjunct.

\begin{defi}[Spatial Family of Hybrid Automata \cite{DammHS15}]
  A \emph{spatial family of hybrid automata (SFHA)} 
is a family of the form
	$S = (\environment, \{S(i) \mid i\in I\})$, where $\{ S(i)
        \mid i \in I \}$ is a system of similar hybrid automata
and $\environment$ is a topology automaton. 
If for every $i \in I$, $S(i)$ is a linear hybrid automaton, we talk
about a \emph{spatial family of linear hybrid automata (SFLHA)}. 
\ignore{
If the topology automaton is timed, we speak of a {\em  spatial family of
  timed (linear) hybrid automata (SFT(L)HA)}.}
An SFLHA $S$ is {\em decoupled} if 
the real-valued variables in the guard of a
mode switch of $S(i)$ can only be reset in a jump by $S(i)$ or by
$\environment$.
\end{defi}

\subsection{Verification} 
The properties of SFLHA we consider here are {\em safety properties}
 of the form: 

\smallskip
~~~~~~~~~~~~~~$ \forall i_1, \dots, i_n \Phi_{\sf safe}(i_1, \dots, i_n).$ 
 
\smallskip 
\noindent Such properties correspond to safety properties with
exhaustive entry conditions considered in \cite{DammHS15} for the case
when $\Phi_{\sf entry} = \top$ and all admissible states $(q, a)$ in a
mode $q$ satisfy the initial conditions,
i.e.\ all states of the systems $S(i)$ are considered to be initial
states. The following result is a specialization of Theorem~1 and
Lemma~2 in \cite{DammHS15} to this special case.

\begin{thm}
A decoupled SFLHA $S = (\environment, \{S(i) \mid i\in I\})$, 
with ${\sf Init}_q(i) = {\sf Inv}_q(i)$ for all $i \in I$ and all $q \in Q(i)$, 
satisfies a safety property $\Phi_{\sf safe}$ for every run
iff the following hold: 
\begin{itemize}
\item[(1)] $\Phi_{\sf safe}$ is preserved under all flows.  
\item[(2)] $\Phi_{\sf safe}$ is preserved under all jumps. 
\item[(3)] $\Phi_{\sf safe}$ is invariant under all jumps in any
  component of $S$. 
\item[(4)] $\Phi_{\sf safe}$ is preserved under all topology updates. 
\end{itemize}  
\end{thm}
In \cite{DammHS15} we proved that all these tasks can be expressed 
as reasoning tasks in chains of theory extensions, and identified 
conditions under which the extensions in these chains were local 
or stably local. In particular, for checking invariance under flows 
in the SFLHA we can use for each system an encoding like that in 
Theorem~\ref{transl-invar-par}. 
We proved that for decoupled (non-parametric) SFLHA and properties 
$\Phi_{\sf safe}$ which can be expressed in the fragment of the theory 
of pointers discussed in Section~\ref{examples} with linear arithmetic as
the theory of scalars, the problem of checking properties
(1)--(4) above is decidable and in NP (cf.\ Theorems 10, 11 and 12 in 
\cite{DammHS15}); and that for such decoupled 
parametric SFLHA and safety properties both verification and
constraint generation are exponential (cf.\ Theorems 15 and 16 in 
\cite{DammHS15}).
\subsection{Examples: Constraint generation}
\label{verif}

In \cite{DammIS11,DammHS15} we used H-PILoT for the verification of 
LHA and SFLHA under the assumption that the coefficients in 
all linear inequalities were concrete constants. 
Since we used as an endprover the version of Z3 available at that
time, checking validity of non-linear
constraints by quantifier elimination was problematic. 
We now illustrate by examples how SEH-PILoT can be used for 
determining constraints on parameters under which universally
quantified safety properties are inductive invariants of a SFLHA, also when 
parametric coefficients are allowed, so a reduction to linear real
arithmetic is not possible.

\begin{ex}
Consider the family of $n$ water tanks described in Example~\ref{ex2} ($n$ is a parameter).
We can describe it as a SFLHA $S = ({\sf Top}, \{ S(i), i \in \{ 1,
\dots, n \})$ as follows: 
For every $i \in \{ 1, \dots, n \}$, $S(i)$ is a linear hybrid
automaton:  

\smallskip
$S(i) = (\{ L(i), {\sf infl}(i), {\sf outfl}(i) \}, Q, {\sf flow}(i), {\sf Inv}(i), {\sf Init}(i),  E, {\sf guard}(i), {\sf
  jump}(i))$

\smallskip
\noindent where $Q = \{ s_1, s_2 \}$, $E = \{ e_1, e_2 \}$, where $e_1
= (s_1, s_2)$ and $e_2 = (s_2, s_1)$;  
\begin{itemize}
\item ${\sf flow}_{s_1}(i) := (\dot{L}(i) {=} {\sf in}(i) {-}
  {\sf out}(i) \wedge \dot{\sf infl}(i) {=} {\sf in}(i) \wedge
 o_{\sf min} \leq \dot{\sf outfl}(i) {=} {\sf out}(i))$, \\
${\sf flow}_{s_2}(i) := (\dot{L}(i) {=} {\sf in}(i) {-}
  {\sf out}(i)  \wedge \dot{\sf infl}(i) {=} {\sf in}(i) \wedge
 \dot{\sf outfl}(i) {=} 0 \wedge \dot{\sf outfl}(i) {=} {\sf out}(i))$
 \\
where, for every $i$, ${\sf in}(i)$
  and ${\sf out}(i)$ and $o_{\sf min}$ are parameters (non-negative);
\item ${\sf Inv}_{s_1}(i) = (L(i)  {\geq} L_a, {\sf outfl}(i) {>} 0),
  {\sf Inv}_{s_2}(i) =  (L(i) {<} L_a, {\sf outfl}(i) {=} 0)$,  \\
where $L_a$ is the alarm level  (a positive parameter);
\item ${\sf Init}_{s_k}(i) = {\sf Inv}_{s_k}(i) \wedge L \leq L_{\sf o}$,
  $k = 1, 2$, 
  where $L_{\sf o}$ is the overflow level;
\item ${\sf guard}_{e_1}(i) = (L(i) < L_a)$, 
 ${\sf guard}_{e_2}(i) = (L(i) \geq L_a)$, and \\
${\sf jump}_{e_1}(i) = {\sf jump}_{e_2}(i) = (L'(i) = L(i) \wedge {\sf
  outfl}' = {\sf outfl} \wedge {\sf infl}' = {\sf infl})$. 
\end{itemize}
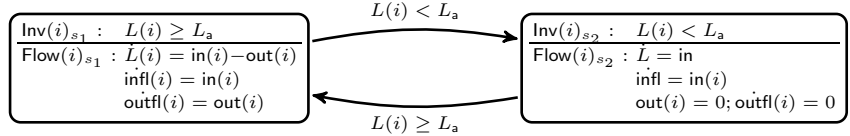
\begin{center}
{\scriptsize 
\vspace{-2mm}\begin{tikzpicture}
\node [square state] (1) {$\begin{array}{ll}
{\sf Inv}(i)_{s_1}: \quad & L(i) \geq L_{\sf a}\\
\hline 
{\sf Flow}(i)_{s_1}: & \dot{L}(i) = {\sf in}(i) {-} {\sf out}(i)\\
& \dot{{\sf infl}}(i) = {\sf in}(i)\\
& \dot{{\sf outfl}}(i) = {\sf out}(i)
\end{array}$}; 
\node [square state] (2) [right = of 1]{$\begin{array}{ll}
{\sf Inv}(i)_{s_2}: \quad  & L(i) <
L_{\sf a}\\
\hline 
{\sf Flow}(i)_{s_2}: & \dot{L} = {\sf in}\\
& \dot{{\sf infl}} = {\sf in}(i)\\
& {\sf out}(i) = 0; \dot{{\sf
    outfl}}(i) = 0
\end{array}$};
 
\path 
(1) edge [bend left=10,above] node  {$L(i) < L_{\sf a}$} (2)
(2) edge [bend left=10,below] node  {$L(i) \geq L_{\sf a}$} (1);  
\end{tikzpicture}
\vspace{-3mm}
}
\end{center}
The connections between systems in ${\sf Top}$ are described by the 
 function $next : \{1, \dots, n{-}1 \} \rightarrow \{ 1, \dots, n
  \}$, $next(i) = i{+}1$ and the constraints ${\sf in}(1) = {\sf in}_0$
and $\forall i (2 \leq i \leq n \rightarrow {\sf in}(i) = {\sf out}(i-1))$.
\noindent 
There are no topology updates. 

\smallskip
\noindent Let $\Phi_{\sf safe} = \forall i (L(i) \leq L_{\sf o})$,
where $L_{\sf o}$
is a parameter representing the overflow level for all water tanks $i$.
The task is to determine relationships between $L_{\sf a}, L_{\sf o},
{\sf in}_0$ and ${\sf out}$ under which 
$\Phi_{\sf safe}$ is guaranteed to be
an inductive invariant of $S$.
\end{ex}
{\bf Verification/Constraint Solving.} To verify that the condition
$\Phi_{\sf safe}$ is an inductive invariant we have to check: 
\begin{itemize}
\item[(1)] The property holds for each system when it is in the
  initial state.

\item[(2)] Invariance under flows. 

\item[(3)] Invariance under jumps. 
\end{itemize}
(1) From the definition of the initial states,
$ \forall i ({\sf Init}_{s_k}(i) \rightarrow L(i) \leq  L_{\sf o})$ is
clearly valid. 

\medskip
\noindent (3) Since the jumps do not change the value of $L$, invariance under jumps
follows immediately. 

\medskip
\noindent (2) 
By Theorem~\ref{transl-invar-par}, $\Phi_{\sf safe}$ is invariant under flows iff the
following conjunctions are unsatisfiable: 

\smallskip
{\small \noindent $\forall i \, ( t_0 {<} t_1 \wedge L(i)(t_0) {\leq} L_o \wedge L(i)(t_0)
{\geq} L_{\sf a}
\wedge  \underline{{\sf flow}}_{s_1}(i)(t_0, t_1) \wedge L(i)(t_1)
{\geq} L_{\sf a} \wedge L(i)(t_1) {>} L_{\sf o})$

\noindent $\forall i \, ( t_0 {<} t_1 \wedge L(i)(t_0) {\leq} L_o \wedge L(i)(t_0) {<}
L_{\sf a}
\wedge \underline{{\sf flow}}_{s_2}(i)(t_0, t_1) \wedge L(i)(t_1) {<}
L_{\sf a} \wedge L(i)(t_1) {>} L_{\sf o})$}

\smallskip
\noindent where $\underline{{\sf flow}}_{s_i}(i) = (L(i)(t_1) - L(i)(t_0)) \leq
({\sf in}(i) - {\sf out}(i))(t_1 - t_0)$, taking into account the
constraints on ${\sf in}$, ${\sf out}$ and ${\sf in}_0$ mentioned
above. 
We present a test in which we used a simplified specification
of the problem: We use the fact that a system $i$ is in state $s_1$
iff $L(i) > L_{\sf a}$ and in state $s_2$ iff $L(i) \leq L_{\sf a}$,
and that in state $s_1$ the outflow rate is positive and has $o_{\sf
  min}$ as a lower bound, and in state $s_2$ the outflow rate is 0. 
This can be expressed by the formulae ${\cal K}_{\sf out}$.

\smallskip
$\forall i \, (1 \leq i \leq n \wedge L(i)(t_0) < L_{\sf a} \rightarrow {\sf
  out}(i) = 0)$

$\forall i \, (1 \leq i \leq n \wedge L(i)(t_0)  \geq L_{\sf a} \rightarrow {\sf
  out}(i) \geq o_{\sf min})$

\smallskip
\noindent The link between the water level at moment $t_0$ and moment $t_1$ is 
expressed by the formula ${\cal K}_{\sf update}$: 

\smallskip
$\forall i \, (1 \leq i \leq n \rightarrow L(i)(t_1) = L(i)(t_0) + ({\sf
  in}(i) - {\sf out}(i))*(t_1 - t_0))$

\smallskip
\noindent The link between input and output is described by the 
formula ${\cal K}_{\sf in}$: 

\smallskip
$\forall i \, (i = 1 \rightarrow {\sf in}(i) = {\sf in}_0)$

$\forall i \, (2 \leq i \leq n \rightarrow {\sf in}(i) = {\sf out}(i-1))$

\smallskip
\noindent We might have additional assumptions ${\cal K}_a$, 
for instance $L_{\sf  a} > 0$, $L_{\sf o} > 0$, possibly also $L_{\sf a} < L_{\sf o}$,
${\sf in}_0 > 0$, $t_0 < t_1$ and  
$\forall i ({\sf out}(i) \geq 0)$. 

\smallskip
\noindent To check whether $\Phi_{\sf safe}$ is invariant under flows
we check that 

\smallskip 
${\cal K}_a \cup {\cal  K}_{\sf out} \cup
{\cal K}_{\sf in} \cup {\cal K}_{\sf update} \wedge \forall i \,(L(i)(t_0)
\leq L_{\sf o}) \models \forall i \,(L(i)(t_1) \leq L_{\sf o})$, i.e.\ 

${\cal K}_a \cup {\cal  K}_{\sf out} \cup
{\cal K}_{\sf in} \cup {\cal K}_{\sf update} \wedge \forall i \,(L(i)(t_0)
\leq L_{\sf o}) \wedge (L(i_0)(t_1) > L_{\sf o}) \models \perp.$

\smallskip
\noindent 
We introduce two function symbols $l$ and $lp$ defined by: 
$l(i) := L(i)(t_0)$ and $lp(i) = L(i)(t_1)$ and adapt    
${\cal K}_{\sf out}$ and ${\cal K}_{\sf update}$
accordingly to ${\cal K}_o$ and ${\cal K}_{\sf u}$: 

 \smallskip
${\cal K}_o = \{ \forall i \,(1 \leq i \leq n \wedge l(i) < L_{\sf a} \rightarrow {\sf
  out}(i) = 0), \\
~~~~~~~~~~~~~~\forall i \,( 1 \leq i \leq n \wedge l(i) \geq L_{\sf a} \rightarrow {\sf
  out}(i) \geq o_{\sf min}) \}$

${\cal K}_u = \{ \forall i \,(1 \leq i \leq n \rightarrow lp(i) = l(i) + ({\sf
  in}(i) - {\sf out}(i))*(t_1 - t_0)) \}$

\smallskip
\noindent 
We can structure the theory axiomatized by ${\cal K}_a \cup {\cal  K}_{\sf out} \cup
{\cal K}_{\sf in} \cup {\cal K}_{\sf u} \cup {\cal K}_l$ (where ${\cal
  K}_l = \{  \forall i (1 \leq i \leq n \rightarrow 0 \leq l(i)
\leq L_{\sf o}) \}$)  as a chain of theory extensions as follows: 
$$\T_0 ~\subseteq~ \T_0 \cup {\cal K}_l ~\subseteq~ \T_0 \cup  {\cal K}_l \cup {\cal K}_o
~\subseteq~ \T_0 \cup  {\cal K}_l \cup {\cal K}_o \cup {\cal K}_{\sf in}  ~\subseteq~ \T_0
\cup {\cal K}_o  \cup  {\cal K}_l \cup {\cal K}_{\sf in}  \cup {\cal K}_u,$$
where $\T_0$ is the disjoint combination of linear integer arithmetic (for the
indices) with the theory ${\mathbb R}$ of real numbers (the theory of
real closed fields). 
\begin{itemize}
\item The extension $\T_0 \subseteq \T_0 \cup {\cal K}_l$ is an extension
of $\T_0$ with the new function symbol $l$ satisfying a boundedness
condition, so by the results in Section~\ref{examples} is local. 
\item The extension $\T_0 \cup {\cal K}_l \subseteq \T_0 \cup  {\cal
    K}_l \cup {\cal K}_o$ is an extension of $\T_0 \cup {\cal K}_l$
  with a new function ${\sf out}$ satisfying the guarded boundedness
  conditions ${\cal K}_o$, so by the results in Section~\ref{examples}
  is local. 
\item The extension $\T_0 \cup  {\cal K}_l \cup {\cal K}_o
\subseteq \T_0 \cup  {\cal K}_l \cup {\cal K}_o \cup {\cal K}_{\sf
  in}$ is an extension of $\T_0 \cup  {\cal K}_l \cup {\cal K}_o$ with
a new function ${\sf in}$ satisfying the axioms ${\cal
  K}_{\sf in}$, so being an extension by definitions is local.
\item The extension $\T_0 \cup  {\cal K}_l \cup {\cal K}_o \cup {\cal K}_{\sf in}  \subseteq \T_0
\cup {\cal K}_o \cup {\cal K}_{\sf in}  \cup {\cal K}_u,$ is an
extension with a new function $lp$ satisfying the definitions ${\cal
  K}_u$, so it also is local.
\end{itemize}
To check whether 
$\T_0 \cup {\cal K}_o \cup {\cal K}_{\sf in}  \cup
{\cal K}_u \cup G \models \perp,$
where $G := \{ 1 \leq i_0, i_0  \leq n,  L(I_0) >
L_{\sf o} \}$, we apply the hierarchical reduction in Theorem~\ref{lemma-rel-transl}
several times, until we reduce the problem to checking the
satisfiability of a ground formula w.r.t.\ $\T_0$. 

We can use H-PILoT to check the satisfiability. H-PILoT performs this 
hierarchical reduction in several steps and detects satisfiability.
The fact that all these extensions are local also allows us to use
Algorithm~1 to derive a universal formula $\Gamma$, representing the 
weakest universally quantified conditions on parameters under which 
unsatisfiability of 
$$\T_0 \cup {\cal K}_o \cup {\cal K}_{\sf in}  \cup
{\cal K}_u \cup \Gamma \cup G,$$
i.e.\ invariance under flows, can be guaranteed.

\medskip
\noindent {\bf Test with SEH-PILoT.} Assume that 
$\{  {\sf in}_0, {\sf out}, o_{\sf min},L_a,L_o,t_0,t_1\}$  are
parameters. 
We use SEH-PILoT to generate constraints on these
parameters under which $\Phi_{\sf safe}$ is an inductive invariant. 
We used $l(i)$ for $L(i)(t_0)$ and $lp(i)$ for $L(i)(t_1)$; we
ignored, as explained above, the 
variables ${\sf infl}$ and ${\sf outfl}$, and used the
consequences of the specification on ${\sf in}, {\sf out}$, e.g.\ 
the fact that ${\sf out}(i) \geq o_{\sf min}$  if $L(i) \geq
L_a$(mode $s_1$) and ${\sf out}(i) = 0$ if  $L(i) < L_a$ (mode $s_2$).

\smallskip
\noindent Some of the properties of $t_0, t_1, o_{\sf min}$ such as
$t_0 < t_1$, ${\sf in}_0 > 0, o_{\sf min} \geq 0$ and $\forall i ({\sf out}(i) \geq 0)$ 
are included as assumptions which are used for simplification. 

{\scriptsize 
\begin{lstlisting}[frame=single]
tasks:
  water-tanks-sat-constraint_slfq:
     mode: GENERATE_CONSTRAINTS
     solver: REDLOG
     options:
         parameter: [in,in0,out,omin,la,lo,n]
         assumptions: [t0<t1,0<in0,0<=omin,"0 <= out(?)",0<la,la<lo]
         slfq_query: true
     specification_type: HPILOT
     specification_theory: REAL_CLOSED_FIELDS
     specification: & spec_water-tanks-sat
      file: |
       Base_functions:={(-,2,0,real),(+,2,0,real),(*,2,0,real)}
       Extension_functions:={(l,1,1),(in,1,3),(out,1,2),(lp,1,4)}
       Relations:={(<,2),(<=,2),(>,2),(>=,2)}
       Constants:={(in0,real),(t0,real),(t1,real),(la, real),(lo,real)}
       Clauses:=
       (FORALL i). i = _1 --> in(i) = in0;
       (FORALL i). _2 <= i, i <= n --> in(i) = out(i-_1);
       (FORALL i)._1 <= i,i <= n,l(i) < la --> out(i) = _0;
       (FORALL i)._1<=i,i<=n,l(i) >= la --> out(i) >= omin;
       (FORALL i)._1<=i,i<=n --> lp(i) = l(i)+((in(i)-out(i))*(t1-t0));
       (FORALL i)._1 <= i, i <= n --> l(i) <= lo; 
       Query := t0 < t1;  _1 <= i0;  i0 <= n; lp(i0) > lo;
\end{lstlisting}
}

\smallskip 
\noindent 
Below is the output of SEH-PILoT (we formatted the output for clarity): 

{\scriptsize 

\begin{lstlisting}[frame=single]
Metadata:
    Date: '2025-04-11 16:57:11'
    Number of Tasks: 1
    Runtime Sum (s): 1.8127
water-tanks-sat-constraint_slfq:
    Result: (FORALL i0). OR(i0 - _1 < _0, i0 - n > _0, 
           AND((((in0 * t0) - (in0 * t1)) - la) + lo >= _0, 
                    i0 - _1 = _0, out(i0) - omin < _0), 
           AND(i0 - _2 >= _0, out(i0) - omin < _0, 
              (((out(i0 - _1)*t0) - (out(i0 - _1)*t1)) - la) + lo >= _0), 
           AND(i0 - _2 >= _0, out(i0 - _1) - omin < _0), 
           AND(i0 - _2 >= _0, out(i0 - _1) - out(i0) <= _0), 
           AND(i0 - _1 = _0, out(i0) - in0 >= _0), 
           AND(out(i0) > _0, out(i0) - omin < _0))
    Runtime (s): 1.8127
    Statistics:
        (step) created subtask:
            time (ms): 0.0795
        (subtask) Eliminate symbols and negate result:
            water-tanks-sat-constraint_slfq_SE:
                (step) constants introduced by H-PILoT:
                    time (ms): 65.2486
                (step) parameter:
                    time (ms): 0.2755
                (step) constants:
                    time (ms): 0.4599
                (step) execute Redlog:
                    time (ms): 1302.3116
                    num_atoms_before_SLFQ_query: '143'
                    num_atoms_after_SLFQ_query: '25'
                (step) Redlog query:
                    time (ms): 0.0091
                (step) simplified with assumptions:
                    time (ms): 443.5679
                    num_atoms_formula_before_assumptions: '25'
                    num_atoms_formula_after_assumptions: '16'
                (step) translated result:
                    time (ms): 0.7075
\end{lstlisting}
}

\begin{ex}
Consider a family of cars on a highway with two lanes.
The formalization of such systems as  SFLHA $S = ({\sf Top}, \{ S(i)
\mid i \in I \})$, where for every index $i \in I$, $S(i)$ is the
hybrid system in Figure~\ref{fig:cars} (cf.\ also 
\cite{DammHS15}). 

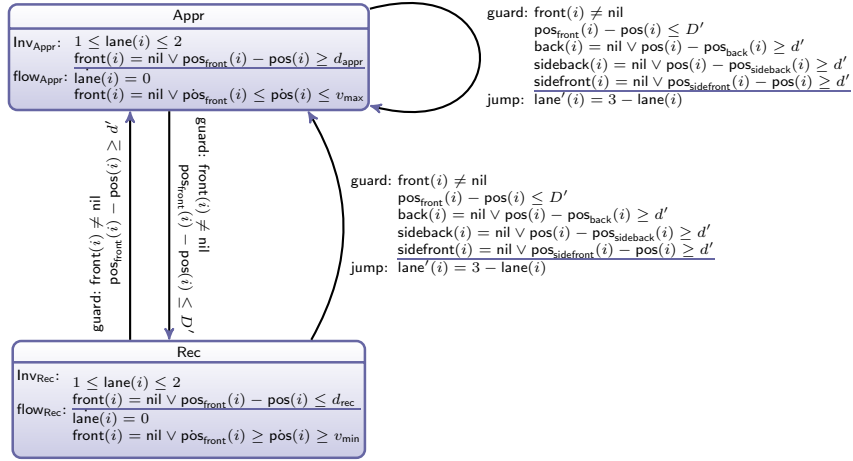
\begin{figure}[h]%
\vspace{-5mm}
\begin{center}
\hspace{-5em}
\arrayrulecolor{blue!40!black!60}
\scalebox{0.8}{
\begin{tikzpicture}
  \HATikzStyle
  \node (accbot) [rectangle split, line width=1.2pt]
             {\textsf{Appr} \nodepart{second} 
    \multimath{{\sf Inv}_{\textsf{Appr}}{:}\\\ \\{\sf flow}_{\textsf{Appr}}{:}\ {}\\\ }%
		\multimath{1\leq \lane{i}\leq 2\\
    \front{i}=\nil\vee {\sf pos}_{\sf front}(i)-\pos{i}\geq d_{\sf appr}\\\hline
    \lane[.]{i}=0\\
		\front{i}=\nil\vee {\dot {\sf pos}}_{\sf front}(i)
                \leq {\pos[.]{i}}\leq v_{\sf max}
		}};
  \node (decbot) [rectangle split, below=3.8cm of accbot, line width=1.2pt]
             {\textsf{Rec} \nodepart{second} 
    \multimath{{\sf Inv}_{\textsf{Rec}}{:}\\\ \\{\sf flow}_{\textsf{Rec}}{:}\ {}\\\ \\\ }%
    \multimath{1\leq \lane{i}\leq 2\\
		\front{i}=\nil\vee {\sf pos}_{\sf
                  front}(i)-\pos{i}\leq d_{\sf rec}\\\hline
    \lane[.]{i}=0\\
    \front{i}=\nil\vee {\dot {\sf pos}}_{\sf front}(i)\geq{\pos[.]{i}}
    \geq v_{\sf min}
		}};
  \tikzstyle{every node}=[font=\scriptsize,fill=none];
  \path[jump,line width=1.2pt]
      (decbot) edge[bend right,transform canvas={xshift=4.5em}]
		          node [right,pos=.5]{
          \multimath{{\sf guard}{:}\ {}\\[4.2em]{\sf jump}{:}}%
					\multimath{
          \front{i}\not=\nil\\ {\sf pos}_{\sf front}(i)-\pos{i}\leq D'
          \\\back{i}=\nil \vee \pos{i}-{\sf pos}_{\sf back}(i)\geq d'
          \\\sideback{i}=\nil \vee \pos{i}-{\sf pos}_{\sf sideback}(i)\geq d'
          \\\sidefront{i}=\nil \vee {\sf pos}_{\sf sidefront}(i)-\pos{i}\geq d'
          \\\hline\lane[']{i}= 3-\lane{i}}}
					(accbot)
      (accbot) edge[loop right, looseness=4]
		          node [right,pos=.5] {
          \multimath{{\sf guard}{:}\ {}\\[4.2em]{\sf jump}{:}}%
					\multimath{
          \front{i}\not=\nil\\ {\sf pos}_{\sf front}(i)-\pos{i}\leq D'
          \\\back{i}=\nil \vee \pos{i}-{\sf pos}_{\sf back}(i)\geq d'
          \\\sideback{i}=\nil \vee \pos{i}-{\sf pos}_{\sf sideback}(i)\geq d'
          \\\sidefront{i}=\nil \vee {\sf pos}_{\sf sidefront}(i)-\pos{i}\geq d'
          \\\hline\lane[']{i}= 3-\lane{i}}}
					(accbot)
     (accbot) edge[transform canvas={xshift=-1em}]
		          node[rotate=-90,anchor=south] {
							                               \multimath{{\sf guard}{:}\ {}\\[1em]}%
							                               \multimath{\front{i}\not=\nil\\ {\sf pos}_{\sf front}(i)-\pos{i}\leq D'
																						}} (decbot)
     (decbot) edge[transform canvas={xshift=-3em}]
		          node[rotate=90,anchor=south] {
							                              \multimath{{\sf guard}{:}\ {}\\[1em]}%
							                              \multimath{\front{i}\not=\nil\\ {\sf pos}_{\sf front}(i)-\pos{i}\geq d'
																						}} (accbot)
        ;
\end{tikzpicture}}
\hspace{-4em}
\vspace{-1em}
\end{center}
\caption{Hybrid automaton modeling the behavior of a car on a two-lane
  highway}%
\label{fig:cars}%
\end{figure} 
\end{ex}
\noindent Let $\Phi_{\sf safe} = \forall i \,({\sf pos}_{\sf front}(i) -
{\sf pos}(i) \geq d_{\sf safe})$, where $d_{\sf safe}$
is a parameter representing the safe distance between a car and the
next car in front of it.
The task is to determine relationships between the parameters 
$v_{\sf min}, v_{\sf max}, d_{\sf appr}, d_{\sf rec}$ and $d_{\sf safe}$ under which 
$\Phi_{\sf safe}$ is guaranteed to be
an inductive invariant of $S$.

\

\noindent {\bf Invariance under flows.} 
By Theorem~\ref{transl-invar-par}, $\Phi_{\sf safe}$ is invariant under flows iff the
following conjunctions are unsatisfiable: 

\medskip
\noindent {\small $t_0 {<} t_1 \wedge \forall i ({\sf pos}_{\sf front}(i)(t_0) {-} {\sf
  pos}(i)(t_0) {\geq} d_{\sf safe}) \wedge  \underline{{\sf flow}}(t_0, t_1)
\wedge  ({\sf pos}_{\sf front}(i_0)(t_1) {-} {\sf
  pos}(i_0)(t_1) {<} d_{\sf safe} )$,}

\medskip 
\noindent {\small \begin{tabular}{@{}l@{}l@{}l}
where $\underline{{\sf flow}}(i) =$ & $\forall i ({\sf Inv}_{\sf appr}(t_0) {\rightarrow}$ & ${\sf pos}_{\sf front}(i)(t_1) {-} {\sf
  pos}(i)(t_1) \leq {\sf pos}_{\sf front}(i)(t_0) {-} {\sf pos}_{\sf
  front}(i)(t_0) \wedge $ \\
& & ${\sf pos}(i)(t_1) \leq {\sf pos}(i)(t_0) + v_{\sf max}*(t_1 - t_0)
\wedge {\sf Inv}_{\sf appr}(t_1)) ~\wedge $\\
& $\forall i ({\sf Inv}_{\sf rec}(t_0) {\rightarrow}$ & ${\sf pos}_{\sf front}(i)(t_1) {-} {\sf
  pos}(i)(t_1) \geq {\sf pos}_{\sf front}(i)(t_0) {-} {\sf pos}_{\sf
  front}(i)(t_0) \wedge $ \\
& & ${\sf pos}(i)(t_1) \geq {\sf pos}(i)(t_0) + v_{\sf min}*(t_1 - t_0)
\wedge {\sf Inv}_{\sf rec}(t_1)). $
\end{tabular}}

\
\noindent We present a test in which we used a slightly simplified description
of the problem, in which the part of the invariant mentioning ${\sf
  lane}(i)$ (which in this case does not change) 
is not included, and it is assumed that ${\sf front}(i) \neq {\sf
  nil}$.

\smallskip
\noindent We used the notation ${\sf p}(i)$ and ${\sf pf}(i)$ for ${\sf pos}(i)(t_0)$  resp.\ ${\sf
  pos}_{\sf front}(i)(t_0)$ and ${\sf pp}(i)$ and ${\sf pfp}(i)$ for ${\sf pos}(i)(t_1)$  resp.\ ${\sf
  pos}_{\sf front}(i)(t_1)$.

\smallskip
To check whether $\Phi_{\sf safe}$ is invariant under flows, we 
need to check that: 
$$\T_0 \cup {\cal K}_{\sf s}  \cup {\cal K}_{\sf u} \cup  G
\models \perp,$$ 
where: 
\begin{itemize}
\item $\T_0$ is the disjoint combination of linear integer arithmetic (for the
indices) with the theory ${\mathbb R}$ of real numbers (the theory of
real closed fields). 
\item ${\cal K}_{\sf s} = \{ \forall i ({\sf pf}(i) - {\sf p}(i) \geq d_{\sf
    safe}) \}$ is the clause form of the safety condition for ${\sf
    pf}, {\sf p}$; 
\item ${\cal K}_{\sf u} = {\cal K}_{\sf pp} \cup {\cal K}_{\sf pfp}
  \cup {\cal K}_{\sf inv}$ is the following set of update axioms: 

$\begin{array}{ll@{}c@{}l}
{\cal K}_{\sf pp} = \{ &   \forall i \,  ({\sf pf}(i)-{\sf p}(i) \geq d_{\sf appr}
& ~\rightarrow ~& {\sf pp}(i)
    \leq {\sf p}(i)+(v_{\sf max}*(t_1-t_0))),\\
     &   \forall i \,  ({\sf pf}(i)-{\sf p}(i) \leq d_{\sf rec} & ~\rightarrow  ~& {\sf pp}(i) 
     \geq {\sf p}(i)+(v_{\sf min}*(t_1-t_0))) \},\\
{\cal K}_{\sf pfp} = \{ & \forall i \, ({\sf pf}(i)-{\sf p}(i) \geq d_{\sf appr} & ~\rightarrow ~&
{\sf pfp}(i)-{\sf pp}(i) \leq {\sf pf}(i)-{\sf p}(i)), \\
        &    \forall i \, ({\sf pf}(i)-{\sf p}(i) \leq d_{\sf rec} & ~\rightarrow ~&
        {\sf pfp}(i)-{\sf pp}(i) \geq {\sf pf}(i) - {\sf p}(i)) \},\\
 {\cal K}_{\sf inv} = \{  &     \forall i \, ({\sf pf}(i)-{\sf p}(i) \geq d_{\sf appr} & ~\rightarrow ~&
        {\sf pfp}(i)-{\sf pp}(i) \geq d_{\sf appr}) \\
      &  \forall i \, ({\sf pf}(i)-{\sf p}(i) \leq d_{\sf rec} & ~\rightarrow ~& 
        {\sf pfp}(i)-{\sf pp}(i) \leq d_{\sf rec}) ~~~ \}
\end{array}$ 
\item $G$ corresponds to the Skolemized negation of $\Phi_{\sf safe}$
  for ${\sf pp}, {\sf pfp}$, i.e.: 

$G = \{ {\sf pfp}(i_0) - {\sf pp}(i_0) < d_{\sf safe} \}$. 
\end{itemize}
We have a chain of theory extensions 
$$ {\cal T}_0 \subseteq {\cal T}_0 \cup {\sf UIF}_{\{ {\sf p} \}} \subseteq
{\cal T}_0 \cup {\cal K}_{\sf s} \subseteq 
 {\cal T}_0 \cup {\cal K}_{\sf s} \cup {\cal K}_{\sf pp}
 \subseteq {\cal T}_0 \cup {\cal K}_{\sf s} \cup {\cal K}_{\sf
   pp} \cup ({\cal K}_{\sf pfp} \cup {\cal K}_{\sf inv})$$
\begin{itemize}
\item The extension $ {\cal T}_0 \subseteq {\cal T}_0 \cup {\sf
    UIF}_{\{ {\sf pp} \}}$ is an extension with a free function symbol, and
  hence local. 
\item The extension ${\cal T}_0 \cup {\sf UIF}_{\{ {\sf pp} \}} \subseteq
{\cal T}_0 \cup {\cal K}_{\sf s}$ is an extension with a function
symbol ${\sf pf}$ axiomatized with a 
boundedness condition $\forall i ({\sf pf}(i) \geq {\sf p}(i) + d_{\sf safe})$
hence it is local. 
\item The extension ${\cal T}_0 \cup {\cal K}_{\sf s} \subseteq 
 {\cal T}_0 \cup {\cal K}_{\sf s} \cup {\cal K}_{\sf pp}$ is an
 extension with a function ${\sf pp}$ axiomatized with boundedness axioms, 
and is therefore local.
\item The extension  ${\cal T}_0 \cup {\cal K}_{\sf s} \cup {\cal K}_{\sf pp}
 \subseteq {\cal T}_0 \cup {\cal K}_{\sf s} \cup {\cal K}_{\sf
   pp} \cup ({\cal K}_{\sf pfp} \cup {\cal K}_{\sf inv})$ is an
 extension with a function ${\sf pfp}$ axiomatized using boundedness axioms,
 and is therefore local.
\end{itemize}

\medskip
\noindent {\bf Tests with SEH-PILoT.} 
Assume that $\{ v_{\sf min}, v_{\sf max}, d_{\sf appr},
d_{\sf rec}, d_{\sf safe} \}$ are parameters. We use SEH-PILoT to generate a set
$\Gamma$ of 
constraints on these parameters under which the system satisfies
condition $\Phi_{\sf safe}$, i.e. 
${\cal T}_0 \cup {\cal  K}_s \cup {\cal K}_u \cup \Gamma \cup G$ is
unsatisfiable. The specification is described below; 
we define the levels of the function symbols ${\sf p}, {\sf pf}, {\sf
  pp}, {\sf pfp}$ in
this extension according to the chain of theory extensions we use: 
${\sf p}$ has level 1, ${\sf pf}$ level 2, ${\sf pp}$ level 3 and
${\sf pfp}$ has level 4. 

\medskip 
{\scriptsize 
\begin{lstlisting}[frame=single]
tasks:
  water-tanks-sat-constraint_slfq:
     mode: GENERATE_CONSTRAINTS
     solver: REDLOG
     options:
         parameter: [vmin,vmax,dappr,drec,dsafe]
         slfq_query: true
     specification_type: HPILOT
     specification_theory: REAL_CLOSED_FIELDS
     specification: &spec_flow-cars-sat
       file: |
        Base_functions := {(-,2,0,real), (+,2,0,real), (*,2,0,real)}
        Extension_functions := {(p,1,1),(pf,1,2),(pp,1,3), (pfp,1,4)}
        Relations := {(<,2),(<=,2),(>,2),(>=,2)}
        Constants := {(t0, real), (t1, real), (vmin, real), (vmax, real), 
                              (dappr, real), (drec, real), (dsafe, real)}
        Clauses :=
        (FORALL i).pf(i)-p(i) >= dappr --> pfp(i)-pp(i) <= pf(i)-p(i);
        (FORALL i).pf(i)-p(i) >= dappr --> pp(i) <= p(i)+(vmax*(t1-t0));
        (FORALL i).pf(i)-p(i) >= dappr --> pfp(i)-pp(i) >= dappr;
        (FORALL i).pf(i)-p(i) <= drec --> pfp(i)-pp(i) >= pf(i) - p(i);
        (FORALL i).pf(i)-p(i) <= drec --> pp(i) >= p(i)+(vmin*(t1-t0));
        (FORALL i).pf(i)-p(i) <= drec --> pfp(i)-pp(i) <= drec;
        (FORALL i).pf(i)-p(i) >= dsafe;
        Query := t0 < t1; 
        % Negated safety condition:
        pfp(i0) - pp(i0) < dsafe;    
\end{lstlisting}
}

\bigskip

\noindent SEH-PILoT returns the following output: 

\medskip
{\scriptsize 
\begin{lstlisting}[frame=single]
Metadata:
    Date: '2025-05-04 23:36:56'
    Number of Tasks: 1
    Runtime Sum (s): 0.3892
test-flow-cars-sat-constraint_slfq:
    Result: AND(dappr - dsafe >= _0, 
                OR(dappr - dsafe = _0, dappr - drec <= _0))
    Runtime (s): 0.3892
    
    Statistics:
        (step) created subtask:
            time (ms): 0.0689
        (subtask) Eliminate symbols and negate result:
            test-flow-cars-sat-constraint_slfq_SE:
                (step) constants introduced by H-PILoT:
                    time (ms): 63.9386
                (step) parameter:
                    time (ms): 0.2474
                (step) constants:
                    time (ms): 0.0634
                (step) execute Redlog:
                    time (ms): 324.6456
                    num_atoms_before_SLFQ_query: '31'
                    num_atoms_after_SLFQ_query: '3'
                (step) Redlog query:
                    time (ms): 0.0087
                (step) translated result:
                    time (ms): 0.2513
\end{lstlisting}
}

\smallskip

\noindent The most time consuming steps are
quantifier elimination and simplification \\(324 ms; the result had 
31 atoms before simplification and 3 after simplification). 

\smallskip
\noindent We also present a version of the test in which $d_{\sf appr}$ and 
$d_{\sf rec}$ depend on the car. We can regard $d_{\sf appr}$ and 
$d_{\sf rec}$ as function symbols introduced in a first theory
extension ${\cal T}_0 \subseteq {\cal T}_1 = {\cal T}_0 \cup {\cal
  K}_d$, so we have the chain of theory extensions: 

\smallskip
$ {\cal T}_0 \subseteq {\cal T}_1 \subseteq {\cal T}_1 \cup {\sf
  UIF}_{\{ {\sf p} \}} \subseteq
{\cal T}_1 \cup {\cal K}_{\sf s} \subseteq 
 {\cal T}_1 \cup {\cal K}_{\sf s} \cup {\cal K}_{\sf pp}
 \subseteq {\cal T}_1 \cup {\cal K}_{\sf s} \cup {\cal K}_{\sf
   pp} \cup ({\cal K}_{\sf pfp} \cup {\cal K}_{\sf inv})$.

\medskip
\noindent We present a test\footnote{We could also consider ${\cal K}_d = \{
  \forall i \,(d_{\sf appr}(i) > 0), \forall i \,(d_{\sf rec}(i) > 0) \}$.}  with SEH-PILoT for
${\cal T}_1 = {\cal T}_0 \cup {\sf UIF}_{\{ d_{\sf
    appr}, d_{\sf rec} \}}$. 

\noindent As before, we consider that 
$\{ v_{\sf min}, v_{\sf max}, d_{\sf appr},
d_{\sf rec}, d_{\sf safe} \}$ are parameters.

\medskip 
{\scriptsize 
\begin{lstlisting}[frame=single]
tasks:
  water-tanks-sat-constraint_slfq:
    mode: GENERATE_CONSTRAINTS
    solver: REDLOG
    options:
        parameter: [vmin,vmax,dappr,drec,dsafe]
        slfq_query: true
    specification_type: HPILOT
    specification_theory: REAL_CLOSED_FIELDS
    specification: &spec_flow-cars-sat
     file: |
      Base_functions := {(-,2,0,real), (+,2,0,real), (*,2,0,real)}
      Extension_functions := {(p,1,2),(pf,1,3),(pp,1,4), (pfp,1,5),
                                             (dappr,1,1),(drec,1,1)}
      Relations := {(<,2),(<=,2),(>,2),(>=,2)}
      Constants := {(t0, real), (t1, real), (vmin, real), (vmax, real), 
                            (dsafe, real)}
      Clauses :=
      (FORALL i).pf(i)-p(i) >= dappr(i) --> pfp(i)-pp(i) <= pf(i)-p(i);
      (FORALL i).pf(i)-p(i) >= dappr(i) --> pp(i) <= p(i)+(vmax*(t1-t0));
      (FORALL i).pf(i)-p(i) >= dappr(i) --> pfp(i)-pp(i) >= dappr(i);
      (FORALL i).pf(i)-p(i) <= drec(i) --> pfp(i)-pp(i) >= pf(i) - p(i);
      (FORALL i).pf(i)-p(i) <= drec(i) --> pp(i) >= p(i)+(vmin*(t1-t0));
      (FORALL i).pf(i)-p(i) <= drec(i) --> pfp(i)-pp(i) <= drec(i);
      (FORALL i).pf(i)-p(i) >= dsafe;
      Query := t0 < t1; 
      % Negated safety condition:
      pfp(i0) - pp(i0) < dsafe;    
\end{lstlisting}
}

\smallskip
\noindent SEH-PILoT returns the following output: 

\smallskip
{\scriptsize 
\begin{lstlisting}[frame=single]
Metadata:
    Date: '2025-05-05 00:00:50'
    Number of Tasks: 1
    Runtime Sum (s): 0.4376
test-flow-cars-sat-constraint_slfq:
    Result: (FORALL i0).AND(dsafe - dappr(i0) <= _0, 
                           OR(dappr(i0)-drec(i0)<=_0,dsafe-dappr(i0)=_0))
    Runtime (s): 0.4376
    Statistics:
        (step) created subtask:
            time (ms): 0.0791
        (subtask) Eliminate symbols and negate result:
            test-flow-cars-sat-constraint_slfq_SE:
                (step) constants introduced by H-PILoT:
                    time (ms): 70.9262
                (step) parameter:
                    time (ms): 0.2745
                (step) constants:
                    time (ms): 0.1161
                (step) execute Redlog:
                    time (ms): 365.8944
                    num_atoms_before_SLFQ_query: '31'
                    num_atoms_after_SLFQ_query: '3'
                (step) Redlog query:
                    time (ms): 0.0083
                (step) translated result:
                    time (ms): 0.3049
\end{lstlisting}
}

\medskip
\noindent {\bf Invariance under jumps: A simplified example.} 
We here only present a simple example: We consider a type of 
jump in system $S(i_0)$ describing a lane change immediately 
following a topology update and followed by an update of the
link to the front car (we restrict to references to 
${\sf sidefront}(i_0), {\sf  front}(i_0)$; a
more complete description can also be analyzed, with similar 
conditions and updates of  ${\sf sideback}(i_0), {\sf
  back}(i_0)$). 
\begin{itemize}
\item The guard of the mode switch is: $ {\sf  pos}_{\sf
    sidefront}(i_0) - {\sf pos}(i_0) > d_{\sf change}$. 

\noindent We assume that the information available to the system is
  correct, i.e.\ \\
${\sf pos}_{\sf sidefront}(i_0) = {\sf pos}({\sf
    sidefront}(i_0))$ and ${\sf pos}_{\sf front}(i_0) = {\sf pos}({\sf
    front}(i_0))$. 

\item The jump condition is: ${\sf front}'(i_0) := {\sf sidefront}(i_0) \wedge {\sf
  sidefront}'(i_0) := {\sf front}(i_0)$.

\item The safety condition is: 
$\Phi_{\sf safe} := \forall i ({\sf pos}({\sf front}(i)) -
  {\sf pos}(i) \geq d_{\sf safe})$. 
\end{itemize}
The task -- for this simplified example -- is to determine the conditions on
$d_{\sf change}$ and $d_{\sf safe}$ under which it is guaranteed that 
after the jump the distance between car $i_0$ and the car in 
front of it is still larger than $d_{\sf safe}$.
This can be reduced to computing a constraint $\Gamma$ on the parameters
under which 
${\cal T}_0 \cup {\cal K}_{\sf safe} \cup G_{\sf update} \cup G_{\sf
  safe} \cup \Gamma$ is unsatisfiable, 
where 

\smallskip
$\begin{array}{lcl}
{\cal K}_{\sf safe} &= \{ & \forall i, j \, ({\sf front}(i) = j \rightarrow {\sf pos}(j) -
  {\sf pos}(i) \geq d_{\sf safe}) \}, \\
G_{\sf safe} & = \{ & {\sf pos}({\sf front}'(t_0)) - {\sf pos}(i_0) < d_{\sf
  safe} \} \text{ and } \\ 
G_{\sf update} & = \{ & {\sf pos}({\sf sidefront}(i_0) - {\sf pos}(i_0) >
d_{\sf change}, \\
& & {\sf front}'(i_0) =  {\sf sidefront}(i_0), {\sf
  sidefront}'(i_0) = {\sf front}(i_0) \}
\end{array}$  

\smallskip 
\noindent We have a chain of local theory extensions: 
${\cal T}_0 \subseteq {\cal T}_0 \cup {\sf UIF}_{\sf pos} \subseteq
{\cal T}_0 \cup {\sf UIF}_{\sf pos} \cup  {\cal K}_{\sf safe}$. 


\medskip
\noindent {\bf Tests with SEH-PILoT.} 
Assume that $\{ d_{\sf change}, d_{\sf safe} \}$ are parameters.
We generate constraints on the
parameters $d_{\sf change}$ and $d_{\sf safe}$ under which $\Phi_{\sf safe}$ is  invariant under this topology update
as follows: 

\medskip
{\scriptsize 
\begin{lstlisting}[frame=single]
tasks:
    lane-change:
        mode: GENERATE_CONSTRAINTS
        solver: REDLOG
        options:
            parameter: [dchange, dsafe]
        specification_type: HPILOT
        specification_theory: REAL_CLOSED_FIELDS
        specification: &spec-lane-change
            file: |
               Base_functions := {(-,2,0,real),(+,2,0,real),(*,2,0,real)}
               Extension_functions := {(front,1,1), (back,1,1), 
                         (sideback,1,1), (sidefront,1,1), (pos,1,2), 
                         (front1,1,2), (back1,1,2), 
                         (sideback1,1,2), (sidefront1,1,2), (pos1,1,3)}
               Relations := {(<,2),(<=,2),(>,2),(>=,2)}
               Constants := {(i0,real), (j0, real), (k0, real),(p0,real),
                         (dsafe,real),(dchange,real)}
               Clauses :=
               (FORALL i, j). front(i)=j --> pos(j)-pos(i) >= dsafe; 
               Query := % Lane change for system i_0
               j0 = front(i0); 
               k0 = sidefront(i0); 
               pos(k0) - pos(i0) > dchange; 
               front1(i0) = sidefront(i0); 
               sidefront1(i0) = front(i0);
               % Negation of the safety property
               p0 = front1(i0);
               pos(p0) - pos(i0) < dsafe;
\end{lstlisting}
}

\medskip 
\noindent 
SEH-PILoT returns the following output: 

\medskip
{\scriptsize 
\begin{lstlisting}[frame=single]
Metadata:
    Date: '2025-02-15 16:16:01'
    Number of Tasks: 1
    Runtime Sum (s): 0.1213
lane-change:
    Result: dchange - dsafe >= _0
    Runtime (s): 0.1213
    Statistics:
        (step) created subtask:
            time (ms): 0.0709
        (subtask) Eliminate symbols and negate result:
            lane-change_SE:
                (step) constants introduced by H-PILoT:
                    time (ms): 67.966
                (step) parameter:
                    time (ms): 0.2401
                (step) constants:
                    time (ms): 0.0533
                (step) execute Redlog:
                    time (ms): 52.5582
                (step) Redlog query:
                    time (ms): 0.0098
                (step) translated result:
                    time (ms): 0.4302
\end{lstlisting}
}

\

\noindent We present a variant of the example, in which $d_{\sf
  change}$ depends on the car, and is modelled as a unary function, 
and in which we added assumptions stating that $\forall i (d_{\sf
  change}(i) \geq 0)$ and $0 \leq d_{\sf safe}$.

\

{\scriptsize 
\begin{lstlisting}[frame=single]
tasks:
    lane-change:
        mode: GENERATE_CONSTRAINTS
        solver: REDLOG
        options:
            parameter: [dchange, dsafe]
            assumptions: ["0 <= dchange(?)", 0 <= dsafe]
        specification_type: HPILOT
        specification_theory: REAL_CLOSED_FIELDS
        specification: &spec-lane-change
            file: |
               Base_functions := {(-,2,0,real),(+,2,0,real),(*,2,0,real)}
               Extension_functions := {(front,1,2), (back,1,2), 
                         (sideback,1,2), (sidefront,1,2), (pos,1,3), 
                         (front1,1,3), (back1,1,3), (dchange,1,1)
                         (sideback1,1,3), (sidefront1,1,3), (pos1,1,4)}
               Relations := {(<,2),(<=,2),(>,2),(>=,2)}
               Constants := {(i0,real), (j0, real), (k0, real),(p0,real),
                         (dsafe,real)}
               Clauses :=
               (FORALL i, j). front(i)=j --> pos(j)-pos(i) >= dsafe; 
               Query := % Lane change for system i_0
               j0 = front(i0); 
               k0 = sidefront(i0); 
               pos(k0) - pos(i0) > dchange(i0); 
               front1(i0) = sidefront(i0); 
               sidefront1(i0) = front(i0);
               % Negation of the safety property
               p0 = front1(i0);
               pos(p0) - pos(i0) < dsafe;
\end{lstlisting}
}

\medskip 
\noindent 
SEH-PILoT returns the following output: 

\medskip
{\scriptsize 
\begin{lstlisting}[frame=single]
Metadata:
    Date: '2025-02-15 16:20:05'
    Number of Tasks: 1
    Runtime Sum (s): 0.3145
lane-change:
    Result: (FORALL i0). dsafe - dchange(i0) <= _0
    Runtime (s): 0.3145
    Statistics:
        (step) created subtask:
            time (ms): 0.0634
        (subtask) Eliminate symbols and negate result:
            lane-change_SE:
                (step) constants introduced by H-PILoT:
                    time (ms): 66.956
                (step) parameter:
                    time (ms): 0.2466
                (step) constants:
                    time (ms): 0.1047
                (step) execute Redlog:
                    time (ms): 53.2632
                (step) Redlog query:
                    time (ms): 0.0096
                (step) simplified with assumptions:
                    time (ms): 193.4116
                    num_atoms_formula_before_assumptions: '3'
                    num_atoms_formula_after_assumptions: '1'
                (step) translated result:
                    time (ms): 0.4096
\end{lstlisting}
}

\section{Conclusions}
\label{Sec:Conclusions}

In this paper we gave an overview of some of our results on the analysis of 
systems of parametric linear hybrid automata, and focused on the
problem of generating constraints on parameters
under which given safety properties are guaranteed to hold. 
We described an implementation of a method for symbol elimination 
that can be used for this, and illustrated its use by means of
examples; the examples we considered so far are parametric versions of
simplified forms of the full specifications of SFLHA which were verified in
\cite{DammHS15}. 
At the moment we cannot perform generation of constraints for the 
invariance properties related to updates of the topology described in
\cite{DammHS15}, 
because for referring to the closest car ahead, behind, etc. we need to use 
formulae with alternations of quantifiers in a theory of pointers, 
a feature which is supported by H-PILoT for verification, but is not yet
supported by SEH-PILoT for constraint generation.

In future work we would like to 
analyze related problems such as invariant strengthening, which 
was studied for systems described by transition constraints in
\cite{PeuterSofronie2019}. We would like 
to better understand the link between existing small model or 
cutoff properties established in the analysis of systems of systems 
and methods we proposed in \cite{Sofronie97,Sofronie-Stokkermans-fct,Sofronie-GETCO09}.

We hope that these results will prove helpful in the analysis of
cyber-physical systems in general, and for the verification of 
automated driving systems in particular -- thus also for the synthesis of automated 
driving systems guaranteed to satisfy given safety properties. 

\

\noindent {\bf Acknowledgments.} The research
reported here was funded by the Deutsche Forschungsgemeinschaft (DFG,
German Research Foundation) – Projektnummer 465447331.

{\bibliographystyle{abbrv}

}

\end{document}